\documentstyle[psfig]{mn}

\def\iron{\hbox{[Fe\,{\sc ii}]}}

\def\hii{\hbox{H\,{\sc ii}}}

\def\he{\hbox{He\,{\sc i}}}
\def\sivi{\hbox{[Si\,{\sc vi}]}}
\def\sivii{\hbox{[Si\,{\sc vii}]}}
\def\caviii{\hbox{[Ca\,{\sc viii}]}}
\def\oiii{\hbox{[O\,{\sc iii}]}}

\def\oiii{\hbox{[O\,{\sc iii}]}}

\def\alix{\hbox{[Al\,{\sc ix}]}}

\def\PM{$\pm$}
\def\kms{km s$^{-1}$}
\def\ccm{cm$^{-3}$}
\def\scm{cm$^{-2}$}
\def\brg{Br$\gamma$}
\def\ha{H$\alpha$}
\def\ak{A_{\rmn{2.2}}}
\def\ergs{ergs cm$^{-2}$ s$^{-1}$}
\def\h2{H$_2$}

\newif\ifAMStwofonts
 \AMStwofontstrue

\ifoldfss
  \newcommand{\rmn}[1] {{\rm #1}}

  \ifCUPmtlplainloaded \else
    \NewTextAlphabet{textbfit} {cmbxti10} {}
    \NewTextAlphabet{textbfss} {cmssbx10} {}
    \NewMathAlphabet{mathbfit} {cmbxti10} {} 
    \NewMathAlphabet{mathbfss} {cmssbx10} {} 
  \fi
  \ifAMStwofonts
    \ifCUPmtlplainloaded \else
      \NewSymbolFont{upmath} {eurm10}
      \NewSymbolFont{AMSa} {msam10}
      \NewMathSymbol{\upi}     {0}{upmath}{19}
      \NewMathSymbol{\umu}     {0}{upmath}{16}
      \NewMathSymbol{\upartial}{0}{upmath}{40}
      \NewMathSymbol{\leqslant}{3}{AMSa}{36}
      \NewMathSymbol{\geqslant}{3}{AMSa}{3E}

      \let\leq=\leqslant 
      \let\geq=\geqslant 
    \fi
  \fi
\fi 

\ifnfssone
  \newmathalphabet{\mathit}
  \addtoversion{normal}{\mathit}{cmr}{m}{it}
  \addtoversion{bold}{\mathit}{cmr}{bx}{it}
  \newcommand{\rmn}[1] {\mathrm{#1}}

  \newmathalphabet{\mathbfit} 
  \addtoversion{normal}{\mathbfit}{cmr}{bx}{it}
  \addtoversion{bold}{\mathbfit}{cmr}{bx}{it}
  \newmathalphabet{\mathbfss} 
  \addtoversion{normal}{\mathbfss}{cmss}{bx}{n}
  \addtoversion{bold}{\mathbfss}{cmss}{bx}{n}
  \ifAMStwofonts
    \ifCUPmtlplainloaded \else
      \UseAMStwoboldmath
      \makeatletter
      \new@mathgroup\upmath@group
      \define@mathgroup\mv@normal\upmath@group{eur}{m}{n}
      \define@mathgroup\mv@bold\upmath@group{eur}{b}{n}
      \edef\UPM{\hexnumber\upmath@group}
      \new@mathgroup\amsa@group
      \define@mathgroup\mv@normal\amsa@group{msa}{m}{n}
      \define@mathgroup\mv@bold\amsa@group{msa}{m}{n}
      \edef\AMSa{\hexnumber\amsa@group}
      \makeatother
      \mathchardef\upi="0\UPM19
      \mathchardef\umu="0\UPM16
      \mathchardef\upartial="0\UPM40
      \mathchardef\leqslant="3\AMSa36
      \mathchardef\geqslant="3\AMSa3E

      \let\leq=\leqslant 
      \let\geq=\geqslant 
    \fi
  \fi
\fi 

\ifnfsstwo
  \newcommand{\rmn}[1] {\mathrm{#1}}

  \DeclareMathAlphabet{\mathbfit}{OT1}{cmr}{bx}{it}
  \SetMathAlphabet\mathbfit{bold}{OT1}{cmr}{bx}{it}
  \DeclareMathAlphabet{\mathbfss}{OT1}{cmss}{bx}{n}
  \SetMathAlphabet\mathbfss{bold}{OT1}{cmss}{bx}{n}
  \ifAMStwofonts
    \ifCUPmtlplainloaded \else
      \DeclareSymbolFont{UPM}{U}{eur}{m}{n}
      \SetSymbolFont{UPM}{bold}{U}{eur}{b}{n}
      \DeclareSymbolFont{AMSa}{U}{msa}{m}{n}
      \DeclareMathSymbol{\upi}{0}{UPM}{"19}
      \DeclareMathSymbol{\umu}{0}{UPM}{"16}
      \DeclareMathSymbol{\upartial}{0}{UPM}{"40}
      \DeclareMathSymbol{\leqslant}{3}{AMSa}{"36}
      \DeclareMathSymbol{\geqslant}{3}{AMSa}{"3E}

      \let\leq=\leqslant 
      \let\geq=\geqslant 
    \fi
  \fi
\fi 

\ifCUPmtlplainloaded \else
  \ifAMStwofonts \else 
    \def\upi{\pi}
    \def\umu{\mu}
    \def\upartial{\partial}
  \fi
\fi

\title[NIR spectroscopy of nearby Seyferts. I.]{Near-infrared spectroscopy of 
nearby Seyfert galaxies. I. First results\thanks{Based on observations 
collected at the European Southern Observatory, La Silla, Chile.}}
\author[J. Reunanen et al.]{J. Reunanen$^1$, J.K. Kotilainen$^1$, M.A. 
Prieto$^2$\\
$^1$ Tuorla Observatory, University of Turku, V\"ais\"al\"antie 20, 
FIN--21500 Piikki\"o, Finland; reunanen@astro.utu.fi, jarkot@astro.utu.fi\\
$^2$ European Southern Observatory, Karl-Schwarzschild-Str. 2, D--85748 
Garching bei M\"unchen, Germany; aprieto@eso.org}
\date{Accepted date +  Received date }
\pagerange{\pageref{firstpage}--\pageref{lastpage}}
\pubyear{2001}

\begin{document}
\label{firstpage}

\maketitle

\begin{abstract}

We present near-infrared 1.5 -- 2.5 $\mu$m moderate resolution long-slit 
spectra of two Seyfert 1 galaxies (NGC 1097 and NGC 1566), three Seyfert 2s 
(NGC 1386, NGC 4945 and NGC 5128) and one Seyfert 1.5 (NGC 3227), both 
parallel to ionization cone or jet and perpendicular to it. The spectra are 
used to study the spatial extent of the line emission, integrated masses of 
excited {\h2} and the excitation mechanisms of interstellar gas. In all three 
Seyfert 2 galaxies, {\iron} is found to be stronger than {\brg} or H$_2$ 1--0 
S(1), while in the Seyfert 1 NGC 1566 and the Seyfert 1.5 NGC 3227 {\brg} is 
the strongest line. Broad Br$\gamma$ originating from the BLR is detected in 
three 
galaxies (NGC 1386, NGC 1566 and NGC 3227), of which NGC 1386 is optically 
classified as Seyfert 2. In these galaxies {\iron} is narrow and may be X-ray 
excited. In all galaxies except in NGC 5128, the spatial FWHM size 
of H$_2$ 1--0 S(1) nuclear core is larger perpendicular to the cone or jet 
than parallel to it, in agreement with AGN unified models. The values of 
nuclear $N_{\rmn{H_2}}$ are higher in Seyfert 2s than in Seyfert 1s, with the 
Seyfert 1.5 NGC 3227 lying between them. The exception to this is the Seyfert 
2 NGC 1386, where two extended regions are detected parallel to cone. Coronal 
lines are detected in two galaxies, NGC 1386 and NGC 3227. 

\end{abstract}


\section{Introduction}

Arguably the most important advance in recent research of active galaxies has 
been the development of unified models (e.g. Antonucci 1993), where a thick 
molecular torus surrounds the nucleus. In Seyfert 1 (Sy1) galaxies the 
nucleus and the Broad Line Region (BLR) are directly visible, while in 
Seyfert 2 (Sy2) galaxies the torus obscures the nucleus and the BLR and only 
the Narrow Line Region (NLR) is visible. These models are supported by the 
detection in many Sy2s of broad lines in polarized emission (e.g. 
Moran et al. 2000), believed to be light from the BLR reflected
into our line-of-sight, and cone-like structures in narrow-band 
{\oiii} images (e.g. 
Mulchaey, Wilson \& Tsvetanov 1996) almost invariably aligned with the radio jets.

Searches for the obscuring molecular material associated with the torus have 
mostly been carried out at millimeter wavelengths (e.g. 
Maiolino et al. 1997), and consequently they generally give only the total 
molecular content in the nucleus. Better spatial resolution, and therefore 
smaller contamination from any circumnuclear star formation, can be achieved 
by observing in the near-infrared (NIR; e.g. 
Veilleux, Goodrich \& Hill 1997), where in addition to the molecular H$_2$ 
lines a wealth of other emission 
and absorption lines are available. However, the NIR spectroscopic studies 
have mostly been made at moderate spatial resolution ($>$1\farcs5). Only 
recently have better resolution NIR studies been carried out (e.g. Storchi-Bergmann et al. 1999,
Winge et al. 2000), but even these studies have only one slit position angle 
(PA) along either the radio axis or the major axis of inner isophotes in 
{\oiii} images, and thus can not derive the geometry of the molecular emission. 

In this paper we present the results of long-slit 1.5--2.5 $\mu$m moderate 
resolution spectroscopy of the first six galaxies in our sample of nearby 
(500 $<$ $v$ $<$ 1500 km s$^{-1}$) Seyfert galaxies of both types with an ionization 
cone and/or jets. The slit was positioned both perpendicular and parallel to 
the ionization cone or to the jet in order to probe the distribution of the molecular 
material and thus the geometry of the torus. Additionally, broad-band 
$JHK'$-images are available for all the galaxies. These data are used to 
trace the spatial distribution, dynamics and excitation of molecular and 
nebular gas, and to determine the mass of hot ($T > 1000$ K) molecular gas in 
the central regions.

In forthcoming publications, we shall discuss the implications from the 
emission line study of all the 14 Seyferts in our sample, and study the  
stellar populations and star forming histories of the sample, based on the 
absorption lines, and comparison with non-Seyfert spirals.

This paper is organized as follows: In Section 2 the observations, data 
reduction and methods used in the analysis are outlined. In Section 3 the 
galaxies are discussed individually, and in Section 4 we present our 
preliminary conclusions. Throughout this paper, $H_0$ = 75 {\kms} Mpc$^{-1}$ 
and $q_0$ = 0.5 are assumed.

\section{Observations, data reduction and methods of analysis}

Six nearby active galaxies were observed in January 1999 with the 3.6 m ESO 
New Technology Telescope (NTT) using the 1024$\times$1024 px SOFI camera 
(Lidman, Cuby \& Vanzi 2000)and a pixel scale 0\farcs29 px$^{-1}$. The best, 
average and worst seeing during the observations were 0\farcs65, 
$\sim$1\farcs0 and 1\farcs35, respectively. In the spectroscopic observations 
the red grism with resolution $R = 980$ and slit width 1\farcs0 (0\farcs6 for 
NGC 1386) were used. The wavelength range covered is 1.5--2.5 $\mu$m and the 
useful slit length $\sim$2{\arcmin}. Broad-band $JHKs$ images were also 
obtained, except for NGC 1097, for which we used the image presented in 
Kotilainen et al. (2000). The properties of the galaxies, redshift, scale, 
inclination, PA of the major axis, morphology, AGN type, the PAs used, total 
integration times, slit width and FWHM seeing, are given in Table 
\ref{obsprop}.

\begin{table*}
 \caption{Observational properties of the galaxies.}
 \begin{tabular}{llllllllllllll}
  \hline
 Galaxy     & z          &Scale    &i&PA& Morphology & Nucleus &PA$_\parallel$&PA$_\perp$&t$_{int}$& Slit width&Seeing\\
          &            &pc/\arcsec&\degr &\degr &    &         &\degr   &\degr&min     &\arcsec      & \arcsec\\ 
\\
  NGC 1097 &0.00425&82 &46 &141 &SB(s)b     &Sy1  &15,54&-36,-75&2$\times$32,2$\times$48&1\farcs0&1\farcs0\\
  NGC 1386 &0.00289&56 &74 &25  &SB0$^+$    &Sy2  &1    &-89    &48,64      &0\farcs6&1\farcs1\\
  NGC 1566 &0.00483&94 &28 &30  &SAB(s)bc   &Sy1  &-42  & 48    &48,64      &1\farcs0&1\farcs0\\
  NGC 3227 &0.00382&74 &56 &158 &SAB(s)a pec&Sy1.5&15   &-75    &48,80      &1\farcs0&1\farcs0\\
  NGC 4945 &0.00194&19 &78 &43  &SBcdsp     &Sy2  &-64  & 26    &32,64      &1\farcs0&0\farcs9\\
  NGC 5128 &0.00179&19 &65 &30  &E2         &Sy2  &50   &-40    &16,48      &1\farcs0&0\farcs9\\
  \hline 
 \end{tabular}
\label{obsprop}
\end{table*}

The long-slit spectra were taken parallel and perpendicular to the PA of the 
cone or jet. The integrations were taken in two positions along the slit 
separated by $\sim$2{\arcmin}
and the pairs were subtracted from each other. The images were then 
flatfielded using dome flats. Bad pixels were masked out and any remaining 
cosmic rays were removed manually. The images were wavelength calibrated 
using either OH night sky lines or Xe arc lamp calibration frames. The 
nucleus was traced by fitting a low order polynomial, a strip was extracted, and 
the strips were averaged to erase the residual continuum and OH lines from 
the sky. Finally, the images were divided by an atmospheric standard star, 
flux calibrated and averaged. {\sc IRAF}\footnote{{\sc IRAF} is distributed 
by the National Optical Astronomy Observatories, which are operated by the 
Association of Universities for Research in Astronomy, Inc., under 
cooperative agreement with the National Science Foundation} was used for all 
stages of the data reduction.

The accurate determination of fluxes and line widths requires a good S/N 
ratio, and therefore two or more pixels were binned together. However, the 
sizes given later for the nuclear sources are based on unbinned data. In order 
to probe the nuclear region as deeply as possible, all the spectra of a given 
galaxy were combined and the central 1\farcs5 (5 px) spectrum was extracted. 
As a result of this and the slit width (1\farcs0 or 0\farcs6), the effective 
aperture has a diameter of 1\farcs4 or 1\farcs1. The fluxes and widths of 
the lines detected in this aperture are given in Table \ref{nucflux}. For few 
lines, especially for {\caviii} in NGC 1386, an outlaying region with no line 
emission was subtracted to remove the underlaying absorption lines. The 
quoted errors in the following sections are generally dominated by the 
continuum fitting, and are 1$\sigma$ except 3$\sigma$ for upper limits.

\subsection{Extinction}
\label{extinction}

We have estimated the extinction by comparing the observed integrated colours 
from the spectra with the colours of normal, unobscured spiral galaxies 
($H$-$K$ = 0.22, Hunt et al. 1997), as the $H$-$K$ color of S0 or E galaxies 
resemble the color of spiral galaxies (Fioc \& Rocca-Volmerange 1999). We note 
that the extinction based on 
continuum colours or the slope of the spectra may not be correct for the line 
emission. However, no useful line pairs were generally detected. Only in 
NGC 4945 the Br$\zeta$ and Br$\eta$ lines are visible in a few regions and 
the extinction based on them is similar to that derived from the continuum 
colours of the spectra. The difficulty of detecting other {\hii} lines arises 
from the high extinction these galaxies suffer, the intrinsic weakness of the 
{\hii} lines and/or the contamination from nearby absorption lines. We have 
assumed the extinction law ${A_\lambda} \propto \lambda^{-1.85}$ 
(Landini et al. 1984)
and a foreground dust screen. The extinction is given in Table \ref{nucflux} 
for the central region and in the lowest panel of the spatial profile figures. 
Our extinction measurements are in fairly good agreement with the previous 
derivations: e.g. in NGC 4945 Marconi et al. (1996) derived $A_{\rmn{V}} \sim 1.1$
from the continuum colors and $A_{\rmn{V}} > 1.3$ in the Pa$\alpha$ ring. The observed 
colors are also in agreement with previous measurements (e.g. $H$-$K$ = 0.34 Kotilainen 
et al. 2000; here 0.37).

In addition to the hydrogen recombination lines, {\iron} lines can be used to 
estimate the extinction. A useful line pair in the 1.5-2.5 $\mu$m range is 
1.5339 and 1.6773 $\mu$m, which originate from the same $^4$D$_{5/2}$ level. 
These faint emission lines are detected in NGC 1386, NGC 1566, NGC 3227 and 
NGC 5128, as marked in the respective figures. However, the lines are generally 
much broader and more extended than the 
main 1.64 $\mu$m line. For this reason, they have not been used in the 
analysis. These features are not artificial or telluric as the velocity field 
derived from them agrees with the general rotation of the galaxies. 

\begin{figure*}
  \begin{minipage}[t]{0.5\linewidth}
    \psfig{figure=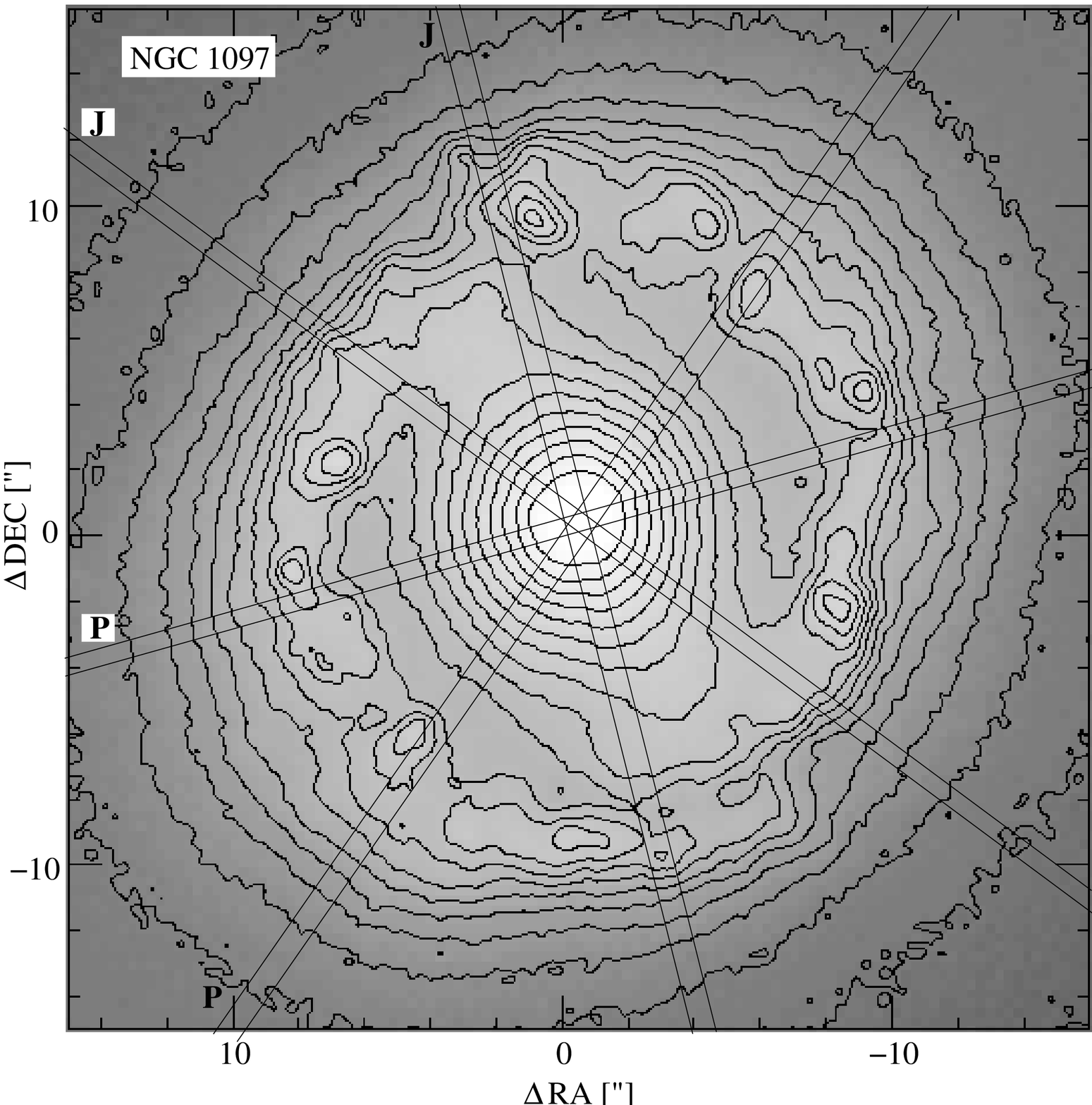,height=7.25cm}
  \end{minipage}
  \begin{minipage}[t]{0.5\linewidth}
    \psfig{figure=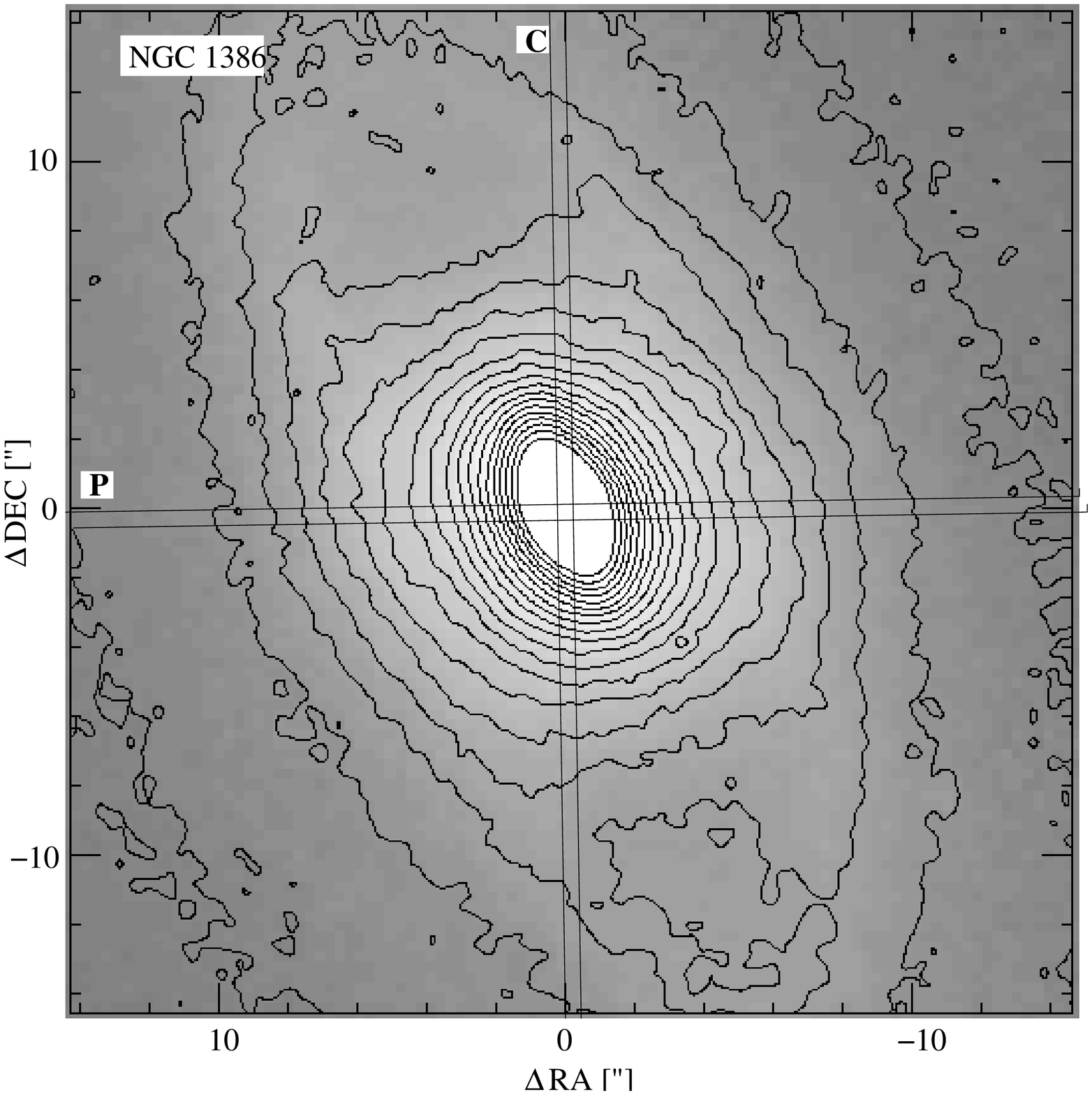,height=7.25cm}
  \end{minipage}

  \begin{minipage}[t]{0.5\linewidth}
    \psfig{figure=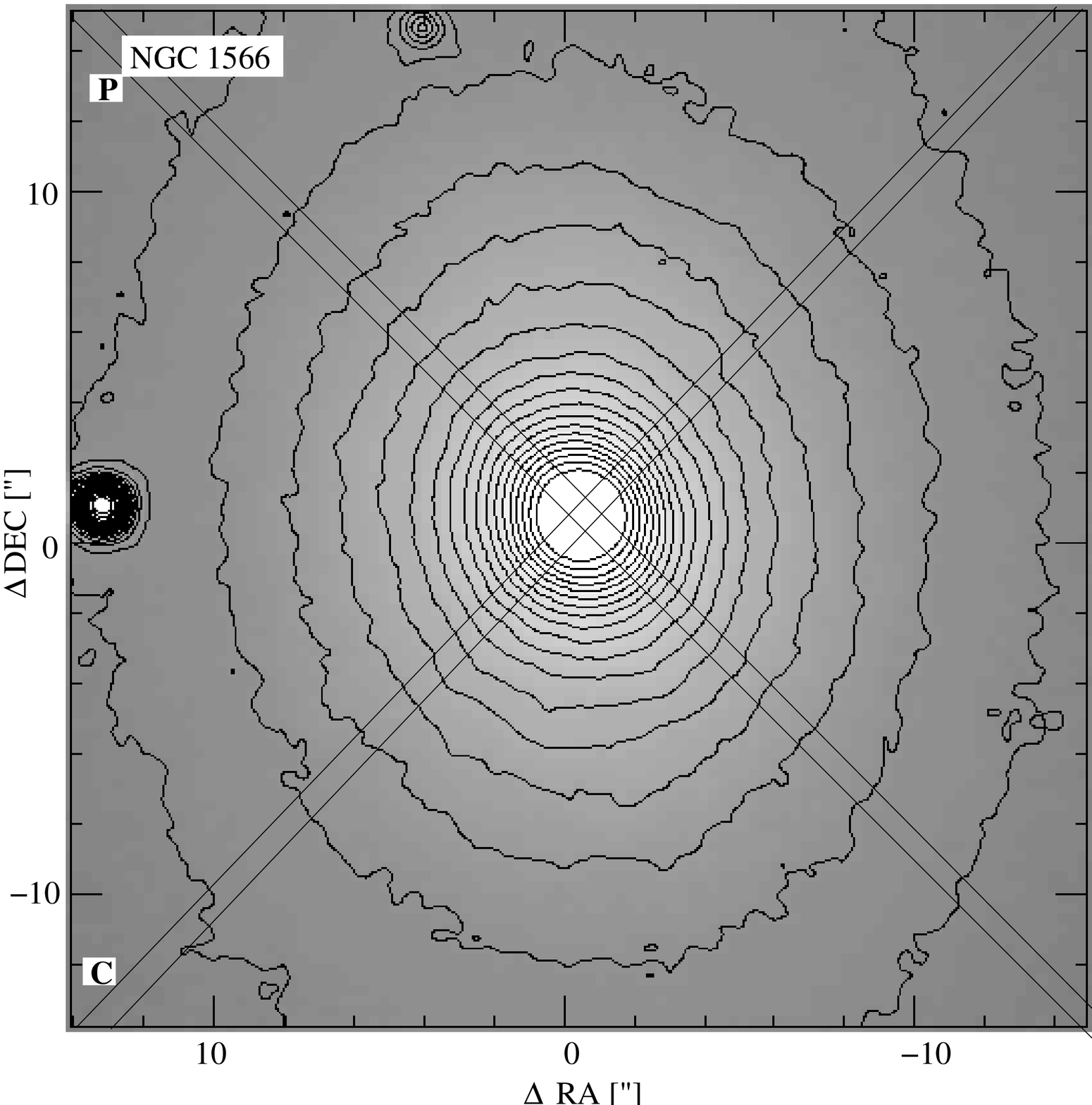,height=7.25cm}
  \end{minipage}
  \begin{minipage}[t]{0.5\linewidth}
    \psfig{figure=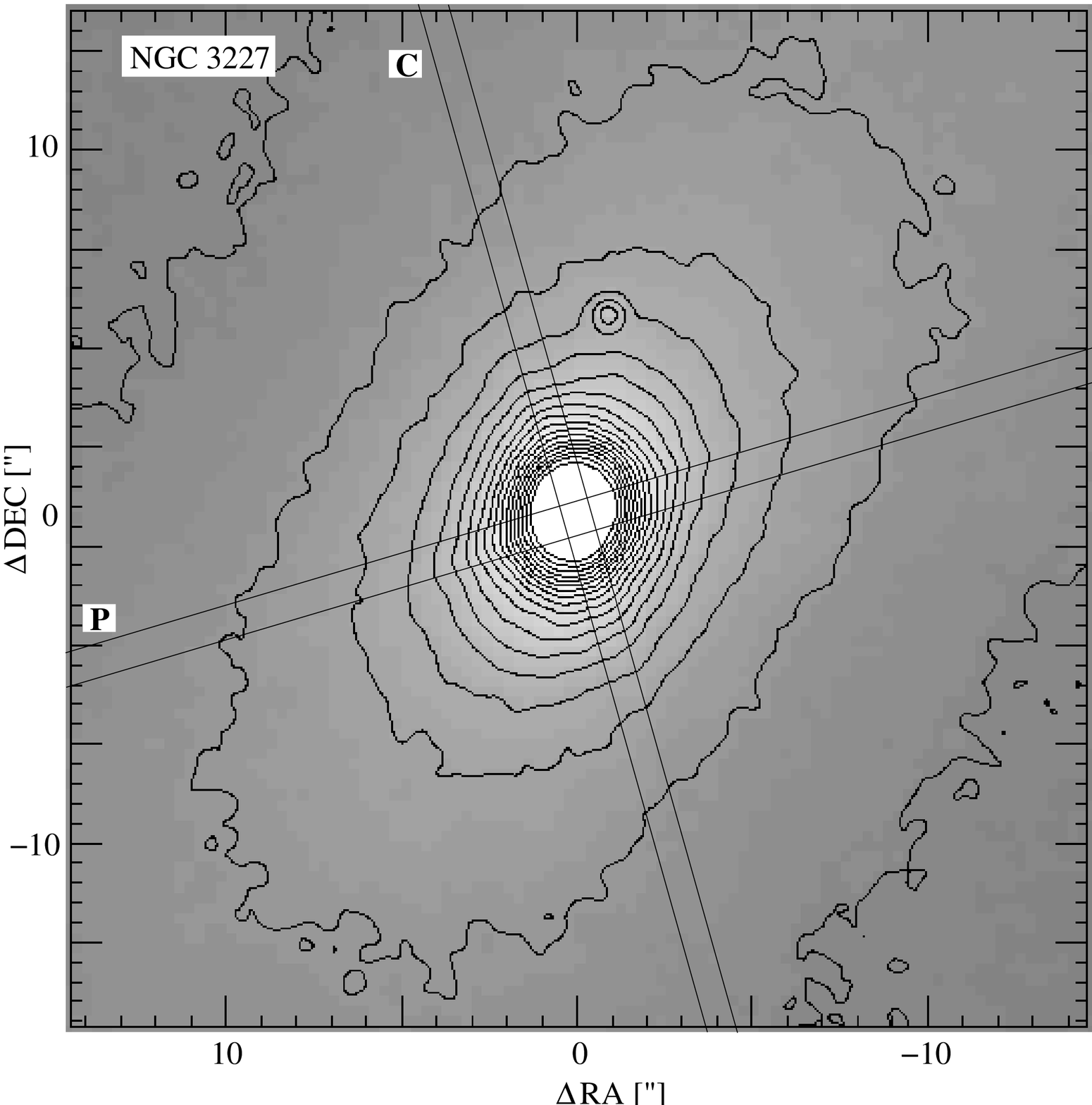,height=7.25cm}
  \end{minipage}

  \begin{minipage}[t]{0.5\linewidth}
   \psfig{figure=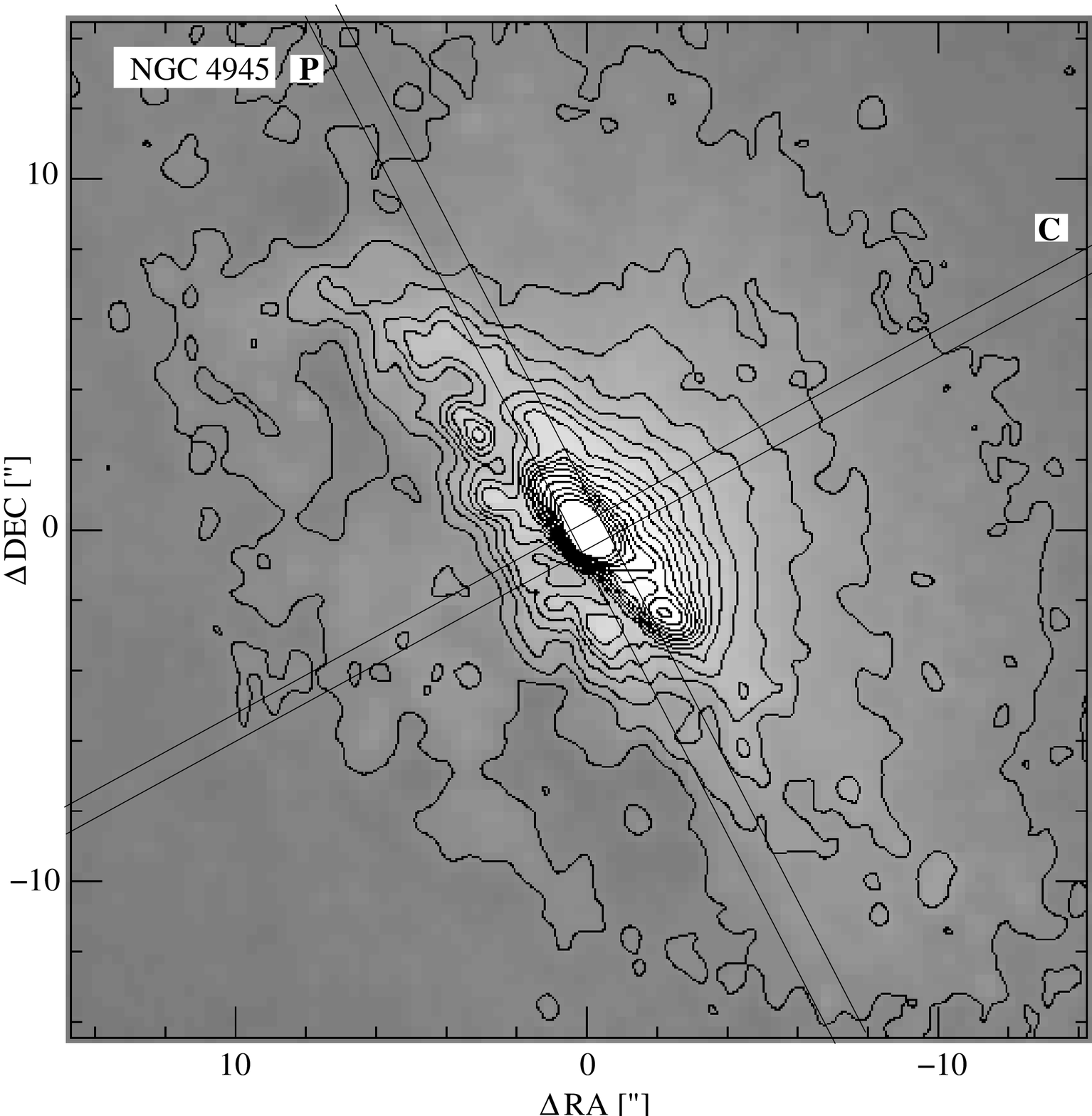,height=7.25cm}
  \end{minipage}
  \begin{minipage}[t]{0.5\linewidth}
   \psfig{figure=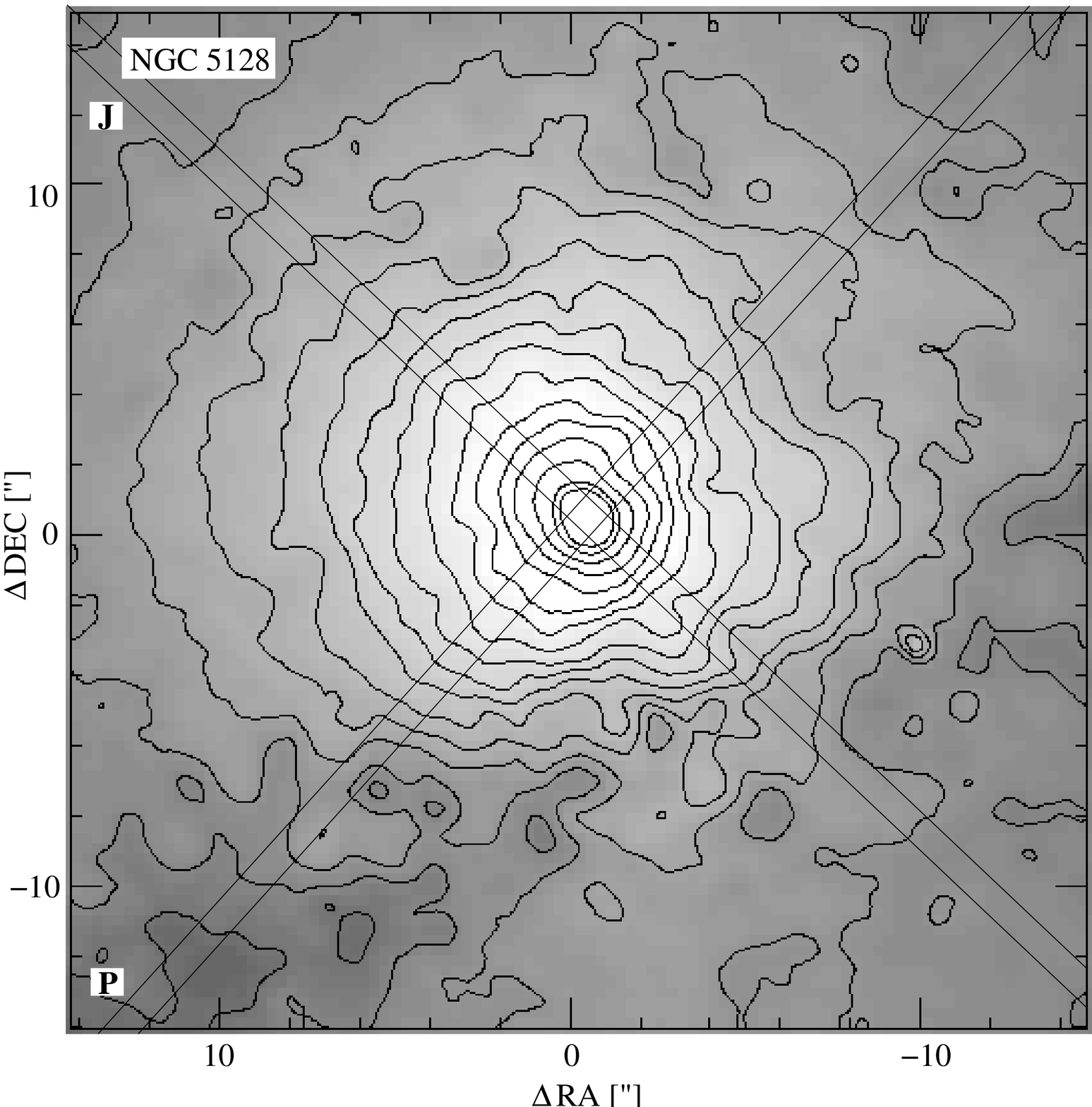,height=7.25cm}
  \end{minipage}
 \caption{The $K$-band image of NGC 1097 ({\em upper left}) from 
Kotilainen et al. (2000), and the $K$$s$-band images of NGC 1386 
({\em upper right}), NGC 1566 ({\em middle left}), NGC 3227 
({\em middle right}), NGC 4945 ({\em lower left}) and NGC 5128 
({\em lower right}). North is up and East to the left. Slit positions are 
indicated in the images, marked with C for parallel to the cone on the side 
where the cone is stronger, J parallel 
to the jet and P perpendicular to them.\label{broad}}
\end{figure*}

\begin{figure*}
 \psfig{figure=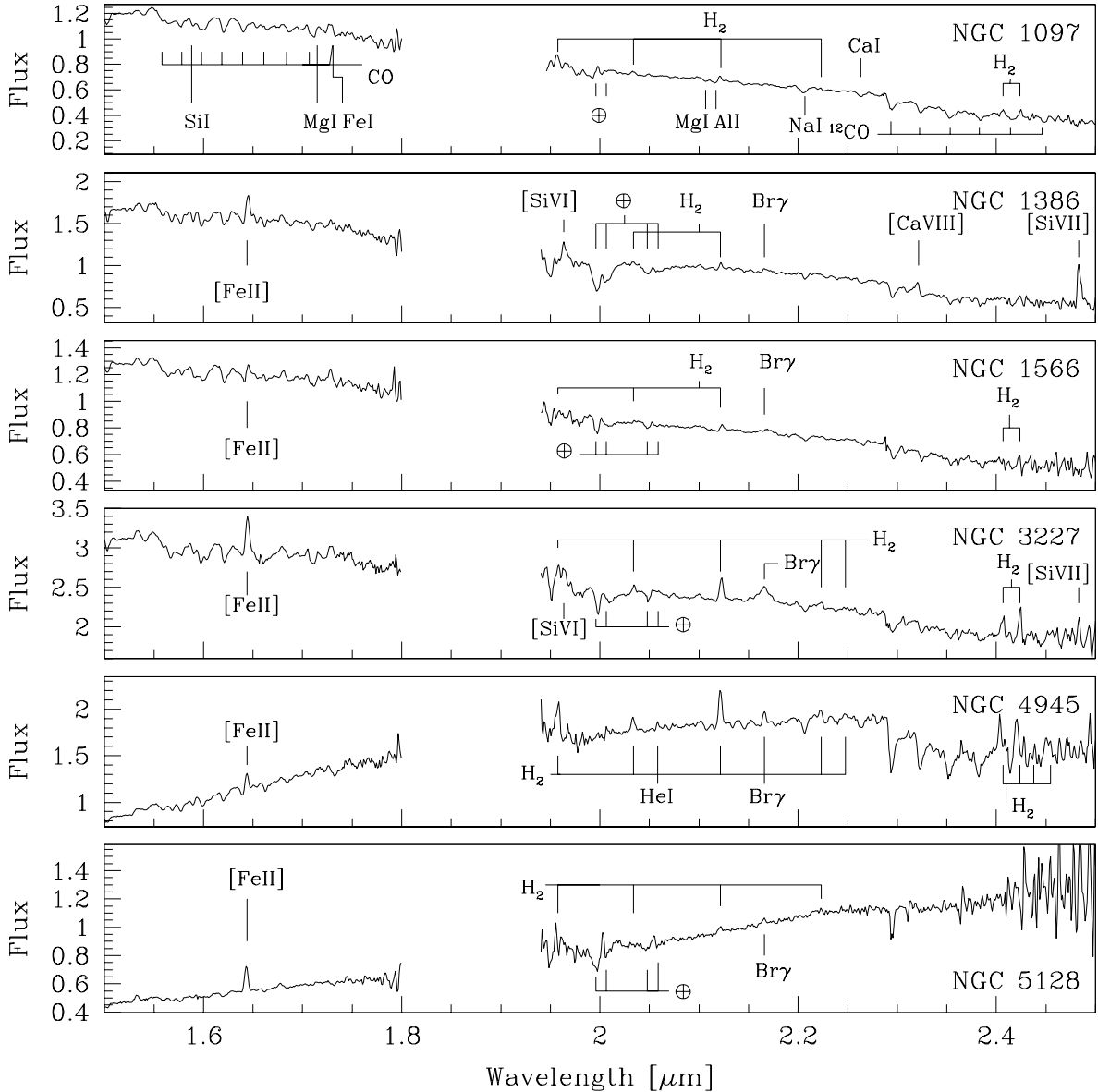,height=17cm}
 \caption{The nuclear $H$- and $K$-band spectra in a 1\farcs4 (1\farcs1 for 
NGC 1386) aperture for, from top to bottom, NGC 1097, NGC 1386, NGC 1566, 
NGC 3227, NGC 4945 and NGC 5128. The flux units are 
10$^{-15}$ ergs s$^{-1}$ cm$^{-2}$ \AA$^{-1}$. 
The main emission lines are marked in each panel, and absorption lines in the 
first panel. Atmospheric features are marked with $\oplus$. \label{spectra}}
\end{figure*}

\begin{figure*}
 \psfig{figure=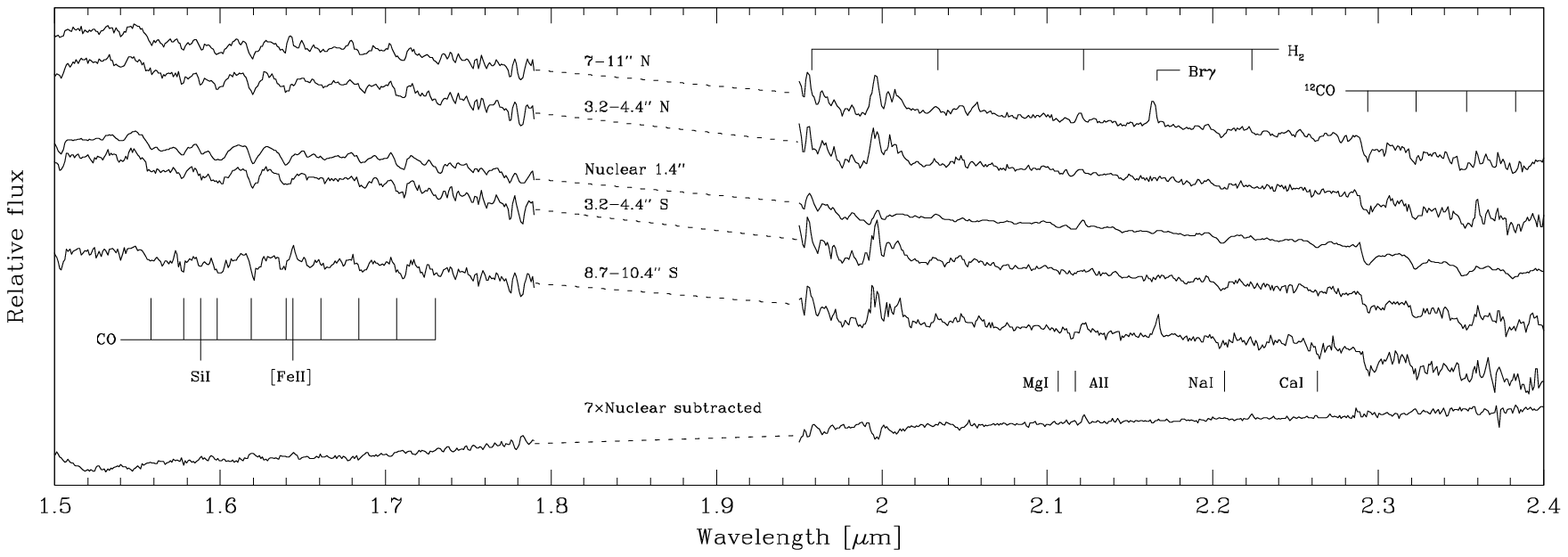}
 \caption{The 1.5--2.5 $\mu$m spectrum of NGC 1097 at PA$=$15\degr (parallel 
to the jet) at different positions along the slit. The K-band flux was scaled 
to match the flux in nuclear 1\farcs4 aperture, and the spectra were placed 
at uniform intervals to facilitate comparison between different regions. The 
absorption features at 
2.0 $\mu$m and 2.05 $\mu$m are due to poor atmospheric cancellation. The 
nuclear subtracted spectrum was obtained as discussed in the text. 
\label{comp1097K}}
\end{figure*}

\subsection{H$_2$ molecular gas: excitation and mass estimate}

The emission line ratios can be used to probe the excitation mechanisms of 
the interstellar matter. Sources of line emission can roughly be divided into 
three classes: cool or warm molecular clouds ({\h2} lines), warm ionized 
interstellar gas ({\hii} lines, {\iron}, {\he}) and very hot, highly ionized 
matter near the centre of the galaxies (coronal lines). The main mechanisms 
for the {\h2} emission are thermal (collisional) heating and UV pumping 
(fluorescence). Thermal excitation, either by shocks (e.g. 
Hollenbach \& McKee 1989), by UV radiation in dense ($n \geq 10^4$ \ccm) 
clouds (Sternberg \& Dalgarno 1989) or by X-rays (e.g. 
Gredel \& Dalgarno 1995), is preferred as the dominant mechanism in most 
extragalactic sources. 
 
The H$_2$ line ratio 2--1 S(1)/1--0 S(1) is commonly used to distinguish between 
thermal ($\sim$0.1--0.2) and UV ($\sim$0.55) excitation in low density 
regions. Interpretation of the H$_2$ 2--1 S(1)/1--0 S(1) ratio is somewhat 
hampered, if X-rays play a significant role in the excitation of the 
molecular gas (Maloney, Hollenbach \& Tielens 1996). Qualitative estimation 
for the importance of X-ray excitation can be derived from the strength of 
{\iron}; X-ray excitation predicts {\iron}/{\brg} up to $\sim$20, while {\hii}
regions have {\iron}/{\brg} $<$ 2.5 (Alonso-Herrero et al. 1997). The ratio 2--1 
S(1)/1--0 S(1) is later used to derive the vibrational excitation temperature 
$T_{\rmn{vib}}$ ($T_{\rmn{vib}} \simeq 5600/\ln(1.355\times 
I_{\rmn{1-0S(1)}}/I_{\rmn{2-1S(1)}})$ assuming Einstein $A$-terms by Turner, 
Kirby-Docken \& Dalgarno 1977), while the rotational temperature is generally 
derived from the line ratio 1--0 S(0)/1--0 S(2) 
($T_{\rmn{rot}} \simeq -1113/\ln(0.323\times I_{\rmn{1-0S(2)}}/I_{\rmn{1-0S(0)}})$).

The equation for the mass of the excited molecular hydrogen can be derived
assuming $T$ = 2000 K, the 1--0 S(1) transition probability $A_{S(1)}$ = 3.47 $\times$ 
10$^{-7}$ s$^{-1}$ (Turner et al. 1977) and the population fraction in 
the $\nu$ = 1, $J$ = 3 level $f_{\nu=1,J=3}$ = 0.0122 (Scoville et al. 1982)
\[
 m_{\rmn{H_2}} \simeq 5.0875\times10^{13}\, D^2\, I_{1-0S(1)}\, 10^{0.4277{\ak}}
\]
where $m_{\rmn{H_2}}$ is the mass of the excited {\h2} in M$_\odot$, $D$ is 
distance in Mpc, $I_{\rmn{1-0S(1)}}$ is the observed flux in 
ergs cm$^{-2}$ s$^{-1}$ and ${\ak}$ is the 2.2 $\mu$m extinction. Integrated 
masses were derived for all PAs (Table \ref{tablederi}). Since the effective 
length of the slit is 
$\sim$2{\arcmin}, the slit covers most of the galaxy and the inclination 
effect is negligible. We have also determined the average surface density and 
the mass of the excited {\h2} in the nucleus. As our nuclear aperture is not 
significantly larger than the seeing, the average surface density can then be 
used to compare the amount of molecular material between Sy1s and Sy2s.

\section{Results and discussion}

The nuclear 30$\times$30{\arcsec} images of the galaxies are displayed in 
Fig. \ref{broad} and the $H$- and $K$-band nuclear spectra in 
Fig. \ref{spectra}. We were unable to completely remove the telluric features 
at $\sim2.0$ $\mu$m and $\sim2.05$ $\mu$m. These signatures are due to {\h2}O 
and are not only temporally but also spatially variable so they can not be 
removed by self-calibration with a distant region in the galaxy. 

\subsection{NGC 1097}

NGC 1097 is a nearby (distance = 17 Mpc) SB(s)b galaxy with a strong bar at 
PA $\sim$137\degr. It interacts with a small elliptical companion toward NW. 
NGC 1097 has an almost circular starburst ring (diameter 1.5-kpc) and a 
nuclear bar, visible both in the NIR continuum (Fig. \ref{broad}) and in 
the 1--0 S(1) line emission (Kotilainen et al. 2000). In the optical, 
NGC 1097 has four faint straight jets (e.g. Wehrle, Keel \& Jones 1997). 
Originally thought to be a LINER (Phillips et al. 1984), currently NGC 1097 
is classified as a Sy1, based on the detection of broad, variable 
double-peaked {\ha} emission (Storchi-Bergmann, Baldwin \& Wilson 1993). The 
radio nucleus of NGC 1097 is compact and weak (Wolstencroft, Tully \& Perley 1984). 
To our knowledge, this is the first $HK$-band spectroscopic study of 
NGC 1097.

\begin{table*}
 \caption{Observed fluxes within the nuclear 1\farcs4 (1\farcs1 for NGC 1386) 
aperture, after adding all the data from different slit PAs. The fluxes 
(first row) are in units of 10$^{-15}$ ergs cm$^{-2}$ s$^{-1}$ and the FWHMs 
(second row) in \AA. A reliable estimate for the 1--0 S(3) emission in 
NGC 3227 is impossible due to telluric residuals.}
 \label{nucflux}
  \begin{tabular}{llllllll}
   \hline
Line   &$\lambda$&NGC 1097     &NGC 1386    &NGC 1566    &NGC 3227    &NGC 4945    &NGC 5128     \\
       &         &Sy1          &Sy2         &Sy1         &Sy1.5       &Sy2         &Sy2   \\
\\
A$_{2.2}$  &     &0.23         &0.26        &0.37        &0.71        &1.82        &1.62        \\
{\iron}  &1.644  &0.57\PM0.14  &10.2\PM0.32 &2.13\PM0.13 &20.8\PM0.28 &6.26\PM0.20 &23.1\PM0.35\\
         &       &32\PM8       &38\PM1      &33\PM2      &41\PM1      &32\PM1      &40\PM1  \\
1--0 S(3)&1.9576 &2.92\PM0.07  &$<$0.4      &0.7\PM0.2   &            &17.0\PM0.26 &7.30\PM0.20 \\
         &       &33\PM1       &            &22\PM3      &            &45\PM1      &22\PM1  \\
{\sivi}  &1.9635 &$<$0.3       &10.1\PM0.18 &$<$0.3      &6.9\PM0.2   &$<$0.5     &$<$0.7      \\
         &       &             &44\PM1      &            &31\PM1      &            &            \\
1--0 S(2)&2.0338 &1.27\PM0.05  &1.38\PM0.11 &0.29\PM0.03 &4.78\PM0.12 &6.30\PM0.10 &$<$0.6      \\
         &       &43\PM2       &30\PM2      &11\PM2      &37\PM1      &40\PM1      &            \\
{\he}    &2.0581 &$<$0.2       &$<$0.3      &$<$0.2      &$<$0.3      &1.54\PM0.09 &$<$0.6    \\
         &       &             &            &            &            &21\PM1      &            \\
1--0 S(1)&2.1218 &1.32\PM0.04  &1.98\PM0.09 &1.53\PM0.03 &11.6\PM0.08 &15.2\PM0.10 &8..41\PM0.19 \\
         &       &30\PM1       &26\PM1      &35\PM1      &36\PM1      &36\PM1      &37\PM1  \\
{\brg}&2.1661    &$<$0.30     &1.76\PM0.14 &5.60\PM0.10 &22.1\PM0.18 &5.23\PM0.10 &4.40\PM0.20 \\
         &       &             &64\PM3      &137\PM5     &125\PM1     &34\PM1      &36\PM2  \\
1--0 S(0)&2.2235 &0.45\PM0.05  &$<$0.3      &$<$0.2      &3.34\PM0.11 &3.00\PM0.09 &1.62\PM0.08 \\
         &       &22\PM2       &            &            &31\PM1      &30\PM1      &17\PM1  \\
2--1 S(1)&2.2477 &$<$0.2       &$<$0.3      &$<$0.2      &0.93\PM0.08 &2.08\PM0.10 &$<$0.52     \\
         &       &             &            &            &29\PM2      &34\PM1      &       \\
{\caviii}&2.3213 &$<$0.2       &3.78\PM0.14 &$<$0.3      &$<$0.3      &$<$0.40     &$<$0.6 \\
         &       &             &43\PM2      &            &            &            &       \\
1--0 Q(1)&2.4066 &1.67\PM0.15  &$<$0.5      &0.92\PM0.22 &10.5\PM0.15 &12.4\PM0.11 &$<$0.9 \\
         &       &34\PM3       &            &11\PM4      &39\PM1      &30\PM1      &       \\
1--0 Q(3)&2.4237 &2.40\PM0.13  &$<$0.5      &2.35\PM0.14 &14.3\PM0.14 &13.5\PM0.11 &$<$0.9 \\
         &       &30\PM2       &            &27\PM2      &30\PM1      &32\PM1      &       \\
{\sivii} &2.4833 &$<$1.1       &17.8\PM0.40 &$<$1.70     &6.1\PM0.3   &$<$0.90     &$<$1.5 \\
         &       &             &38\PM1      &            &20\PM1      &            &       \\
 \hline
  \end{tabular}
\end{table*}

\begin{table*}
\caption{The column density of the excited {\h2}, line ratios, the integrated {\h2} 
mass, the nuclear spatial extent and the total extent of the lines for 
the sample galaxies, both parallel to the cone/jet ($\parallel$) and 
perpendicular to it ($\perp$). The spatial extent (FWHM) is measured from unbinned 
data in the plane of the sky and is corrected for instrumental resolution. 
The total extent is given in the plane of the galaxy. M$_{H_2}$ and extent in NGC 
1097 are given within the ring.
\label{tablederi} 
}
  \begin{tabular}{lllllll}
   \hline
Galaxy     &NGC 1097   &NGC 1386   &NGC 1566   &NGC 3227 &NGC 4945 &NGC 5128       \\
Nucleus                            & Sy1    & Sy2    & Sy1    & Sy1.5&Sy2  &Sy2    \\
 Aperture      &1\farcs4&1\farcs1  &1\farcs4&1\farcs4&1\farcs4&1\farcs4            \\
\\                                                                    
N$_{\rmn{H_2}}$ [10$^{17}$ {\scm}] &1.5    &3.7      &1.9     &22.3  &100  &44     \\
M$_{H_2}$       [M$_\odot$]        &25     &17       &42      &280   &70   &32     \\
{\iron}/1--0 S(1)                  &0.50    &6.1     &1.5     &2.8   &1.20 &7.3    \\
{\brg}/1--0 S(1)                   & ...    &0.88    &2.7     &1.9  &0.32 &0.49   \\
1--0 S(0)/1--0 S(1)                &0.33    &$<$0.15 &$<$0.15 &0.27  &0.17 &0.17   \\
1--0 S(2)/1--0 S(1)                &0.98    &0.72    &0.20    &0.44  &0.49 &$<$0.09\\  
2--1 S(1)/1--0 S(1)                &$<$0.15   &$<$0.15 &$<$0.15 &0.08  &0.12 &$<$0.06\\
M$_{H_2}$ $\perp$ [M$_\odot$]      &60      &30      &110     &340   &190  &30     \\
M$_{H_2}$ $\parallel$[M$_\odot$]   &90      &50      &90      &300   &90   &30     \\
T$_{\rmn{vib}}$  [K]               &$<$2550 &$<$2500 &$<$2700 &1950  &2200 &$<$1800\\
FWHM {1--0 S(1)} $\perp$  [pc]     &108     &39      &82      &84    &57   &35     \\
FWHM {1--0 S(1)} $\parallel$  [pc] & 99     &...     &65      &58    &28   &42     \\
FWHM {\iron} $\perp$  [pc]         &...     &...     &...     &43    &50   &...    \\
FWHM {\iron} $\parallel$  [pc]     &...     &...     &...     &51    &50   &...    \\
Extent 1--0 S(1) $\perp$  [pc]     &80      &190     &280     &860   &1160 &720    \\
Extent 1--0 S(1) $\parallel$  [pc] &340     &410     &210     &600   &1560 &240    \\
\hline
\end{tabular}
\end{table*}

\begin{figure}
 \psfig{figure=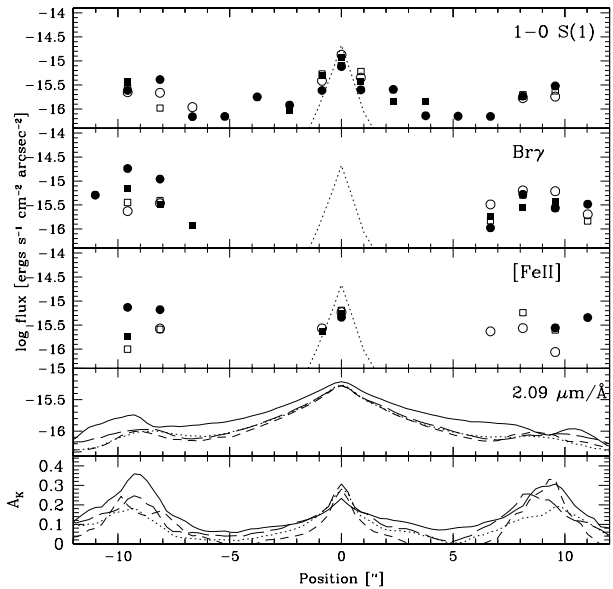,height=8.5cm}
 \caption{The observed line emission in NGC 1097. Open symbols indicate 
perpendicular to the jets (PA of squares = -36\degr, PA of circles = -75\degr) and 
filled symbols parallel to the jets. The PSF is indicated by dotted lines. The 
2.09 $\mu$m continuum emission 
(fourth panel) and extinction (fifth panel) are shown for the PAs of 15\degr 
(solid line), 54\degr (dotted line), -75\degr (short-dashed line) and 
-36\degr (long-dashed line). The positive locations along the slit are in S 
(PA=15\degr), SW (54\degr), SE (-36\degr) and E (-75\degr).\label{n1097jetperp}}
\end{figure}

\begin{figure}
 \psfig{figure=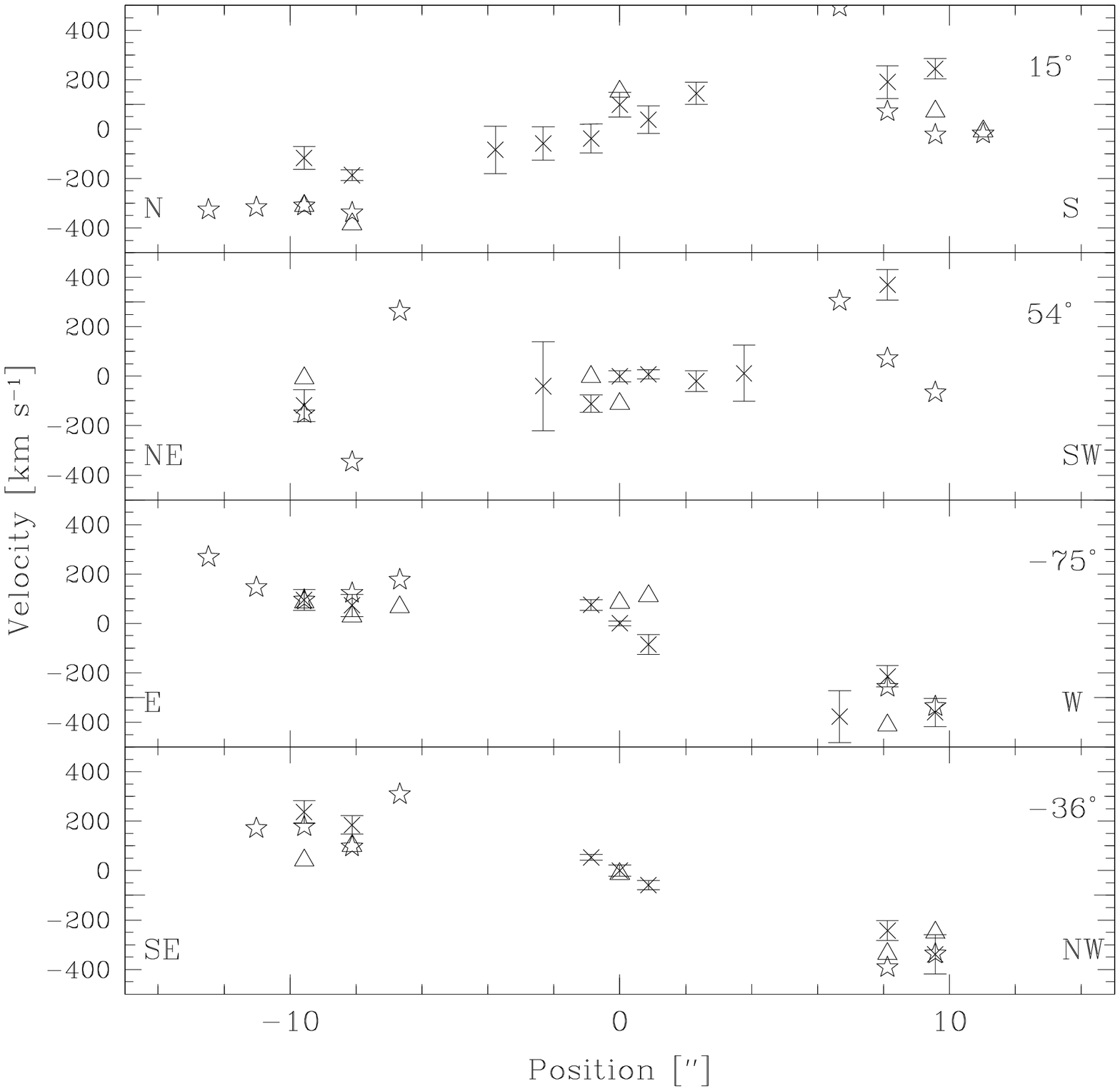,width=8.5cm}
 \caption{The velocity field of NGC 1097 in the most important emission 
lines in the units of km s$^{-1}$, 1--0 S(1) (crosses), {\brg} (stars) and {\iron} 
(triangles).
\label{n1097vel}}
\end{figure}

The $H$- and $K$-band spectra of NGC 1097 at PA 15{\degr} are shown in 
Fig. \ref{comp1097K}. The nuclear subtracted spectrum shown has been obtained 
by matching the flux of the $K$-band CO lines in the off-nuclear 
(0\farcs7-1\farcs6 from the nucleus) spectra to 1\farcs4 nuclear aperture and 
subtracting the result from the nuclear spectra. 
The subtracted spectrum is included here to better show the weak nuclear emission 
lines, especially in the case of NGC 1386 discussed later. The full discussion 
and analysis of the subtracted spectra are postponed to forthcoming papers. 
The only emission lines directly detected in the nucleus are 
the various {\h2} lines (Table \ref{nucflux}). This agrees with the results 
of Kotilainen et al. (2000), who detected nuclear 1--0 S(1) but no {\brg} 
emission in their narrow-band images, suggesting that thermal UV heating is 
not important. The 1--0 S(1) emission is detected further away from the 
nucleus parallel to jets than perpendicular to them (see 
Fig. \ref{n1097jetperp}). Parallel to jets it can be followed up to 
2--3{\arcsec} ($\sim$280 pc in the plane of the galaxy) from the nucleus 
while perpendicular to them only up to $\sim$1{\arcsec} ($\sim$90 pc) 
(Table \ref{tablederi}), in agreement with the 1--0 S(1) image of Kotilainen 
et al. (2000). The distances quoted are corrected for inclination taking 
into account the position angle of the slit unless otherwise stated.

In the $H$-band (Fig. \ref{comp1097K}), {\iron} can only be detected after 
subtracting the underlying CO features. The {\iron} line is narrow (intrinsic 
FWHM 400 \kms), although wider than in typical starbursts ($\la300$ \kms; 
e.g. Moorwood \& Oliva 1988). The 1--0 S(1) line is narrower than {\iron}, 
with intrinsic FWHM 270 \kms. No 2--1 S(1) emission is detected, with 
2--1 S(1)/1--0 S(1) $<$ 0.13 (assuming $T_{\rmn{vib}} < 2400$ K) in the 
central 1\farcs4 aperture, indicating that fluorescent excitation cannot be 
significant. We conclude that in the nucleus there is no current star 
formation and the {\h2} gas is collisionally excited. 

The derived parameters, including the density and mass of molecular hydrogen 
and the spatial extent of different lines in NGC 1097 are given in 
Table \ref{tablederi}. In the nuclear 1\farcs4 aperture, the column density of 
excited molecular hydrogen is 1.5$\times$10$^{17}$ cm$^{-2}$ (corresponding to 
$M_{\rmn{H_2}} = 25$ M$_\odot$), assuming $T$ = 2000 K. The integrated mass 
of excited {\h2} inside the starburst ring is larger parallel to the jets 
(40 and 50 M$_\odot$) than perpendicular to them ($\sim$30 M$_\odot$). This 
suggests that {\h2} emission is produced by shocks in the jets. However, 
further away in the galaxy {\h2} emission was not detected, while the optical 
jets continue much further. This discrepancy is in agreement with the 
non-detection of optical emission lines in the jets (Wehrle et al. 1997).

The kinematics of NGC 1097 has been discussed in detail by 
Storchi-Bergmann, Wilson \& Baldwin (1996a), who found the starburst ring to 
be located between the inner Lindblad resonances. Our NIR velocity fields 
(Fig. \ref{n1097vel}) are in good agreement with their {\ha} velocity field. 
There are no significant differences between the kinematics of {\h2}, {\brg} 
or {\iron}. 

\subsection{NGC 1386}

NGC 1386 is a nearby (distance = 11.5 Mpc) SB0$^+$ galaxy in the Fornax 
cluster with a Sy2 nucleus and dust lanes across the nuclear region (e.g. 
Malkan, Gorjian \& Tam 1998). NGC 1386 has a star forming ring with major 
axis diameter $\sim$1.3-kpc (Weaver, Wilson \& Baldwin 1991). The nuclear 
radio core (Nagar et al. 1999) is extended in  PA$\sim170\degr$, which is 
different from the optical major axis PA ($\sim25$\degr). 
Weaver et al. (1991) detected an outflow directed along the major axis. 
Low-resolution ($R$ = 250) $JHK$-spectra were taken by Winge et al. (2000). 
The {\oiii} image of Storchi-Bergmann et al. (1996b) has a separate emission
 region to N. The slit was aligned to include this region.

\begin{figure*}
 \psfig{figure=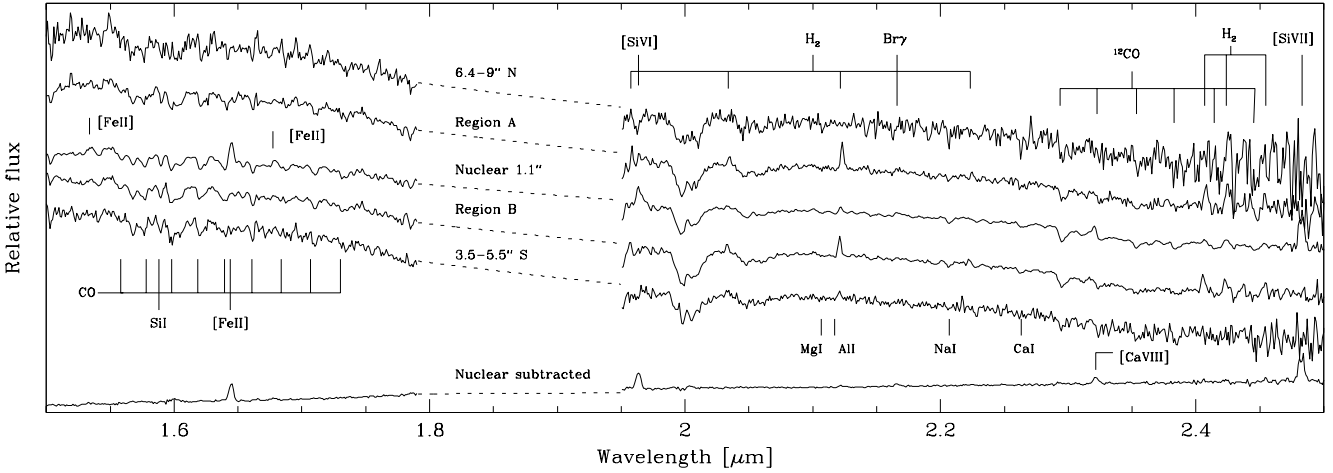}
 \caption{The 1.5 -- 2.5 $\mu$m spectra of NGC 1386 parallel to the cone.  
\label{comp1386K}}
\end{figure*}

\begin{figure}
 \psfig{figure=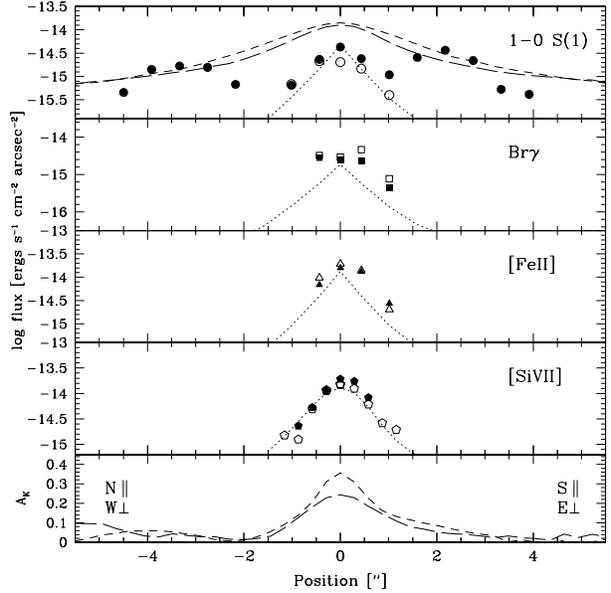,height=8.5cm}
 \caption{The observed line emission in NGC 1386 perpendicular to the 
ionization cone (open symbols) and parallel to it (filled symbols). The PSF 
(dotted line) and the 2.09 $\mu$m continuum emission multiplied by 10 are 
also shown. In the lowest panel the extinction is shown parallel to the cone 
(short-dashed line) and perpendicular to it (long-dashed line).
\label{n1386both}}
\end{figure}

The $H$- and $K$-band spectra of NGC 1386 along the cone are shown in 
Fig. \ref{comp1386K}, while the extent of the lines is displayed in 
Fig. \ref{n1386both}. While {\brg} and {\iron} are only marginally resolved, 
the 1--0 S(1) emission is extended parallel to the cone, where the two 
clearly detected separate regions are discussed in detail below. The {\h2} 
emission can be traced $\sim$4\farcs5 (440 pc in the plane of the galaxy) 
towards N and $\sim$3\farcs5 
(340 pc) towards S. The nuclear 1--0 S(1) core is not resolved parallel 
to the cone, while in the perpendicular direction it is marginally 
resolved. There is no extended emission perpendicular to the cone. 

Three spatially unresolved coronal lines ({\sivi} 1.9635 $\mu$m, 
{\caviii} 2.3213 $\mu$m and {\sivii} 2.4833 $\mu$m) are detected in the 
nucleus of NGC 1386 (Fig. \ref{comp1386K}). Strongest of these coronal lines 
is {\sivii} and it is also overall the strongest observed line in the nuclear 
spectrum. Previously, the {\caviii} coronal line has only been detected in 
the Sy2s Circinus (Oliva et al. 1994) and NGC 1068 (Marconi et al. 1996), and 
unlike in them, {\caviii} in NGC 1386 is already clearly visible before the 
underlying CO features have been subtracted (Fig. \ref{comp1386K}). There is 
a red wing in both {\sivii} and {\sivi}, and possibly in {\caviii}, after 
the CO absorption features have been subtracted. The {\sivi}/{\sivii} ratio in 
the nucleus is 0.48, compatible either with shocks with relatively 
little contribution from UV excitation, or with strong UV continuum 
excitation (Contini \& Viegas 2001). 

We did not detect 2--1 S(1) nor 1--0 S(0) emission in the nucleus. The upper 
limit to 2--1 S(1)/1--0 S(1) ratio, $\leq$ 0.12, is compatible with thermal 
excitation with $T_{\rmn{vib}} \leq 2310$ K. Assuming $T_{\rmn{vib}} = 
2000$ K, we derive 
$N_{\rmn{H_2}}=3.9\times10^{17}$ cm$^{-2}$ (corresponding to 
$M_{\rmn{H_2}} \simeq 20$ M$_\odot$) within the nuclear 1\farcs1 aperture. 
The 1--0 S(1)/{\brg} ratio is 1.14, strongly suggesting that thermal UV 
heating is insignificant, as then {\brg} should be much stronger. On the 
other hand, the \iron/{\brg} ratio is 7.1, suggesting a significant 
contribution from X-rays to the ionization. Since the {\brg} is broad, the 
ratios derived above are in fact lower limits. The likely excitation mechanism 
for nebular gas is therefore X-rays, further supported by the much larger 
width of {\iron} (560 \kms) and {\brg} ($\sim$800 \kms) compared to 1--0 S(1) (190 \kms). 

\begin{figure}
 \psfig{figure=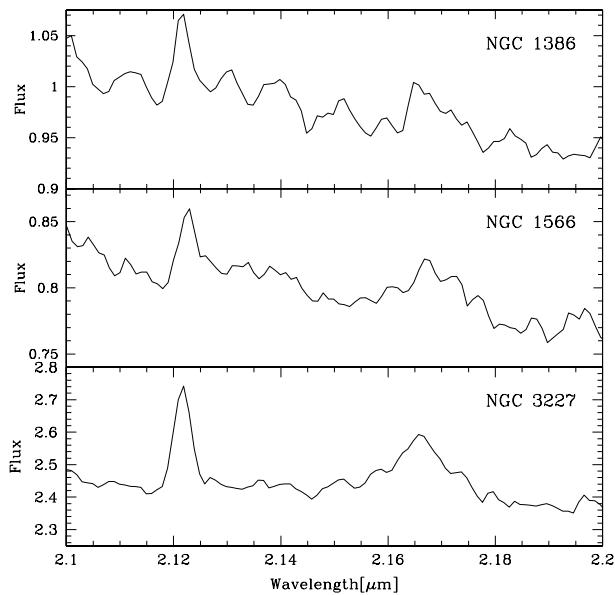,height=8.5cm}
 \caption{Enlarged spectral area around 2.15 $\mu$m showing broad Br$\gamma$ 
and narrow 1--0 S(1) for NGC 1386, NGC 1566 and NGC 3227 from top to bottom, 
respectively, in 1\farcs4 (1\farcs1 for NGC 1386) aperture in units of 
10$^{-15}$ ergs s$^{-1}$ cm$^{-2}$ \AA$^{-1}$}
\label{compbrg}
\end{figure}

The {\brg} line is broad and asymmetric (Fig. \ref{compbrg}),
while both 1--0 S(1) and {\iron} emission are symmetric. {\brg} is broader 
than the optical lines (e.g. Rossa, Dietrich \& Wagner 2000), and is likely 
to originate from the BLR. There is no corresponding broad component to the 
{\iron} line, so the observed {\iron} emission arises outside the BLR. 
{\brg} is concentrated in the nucleus, and the strong {\ha} peak 
(Rossa et al.) is not detected in {\brg}, demonstrating the reduced effect of 
extinction in the NIR. There is no detectable emission from the starburst 
ring, which is not surprising considering its faintness (only a few \% of the 
brightness of the nucleus in {\ha}; Storchi-Bergmann et al. 1996b). However, 
the ring is visible in the $K'$-band (Fig. \ref{broad}). 

\begin{figure}
 \psfig{figure=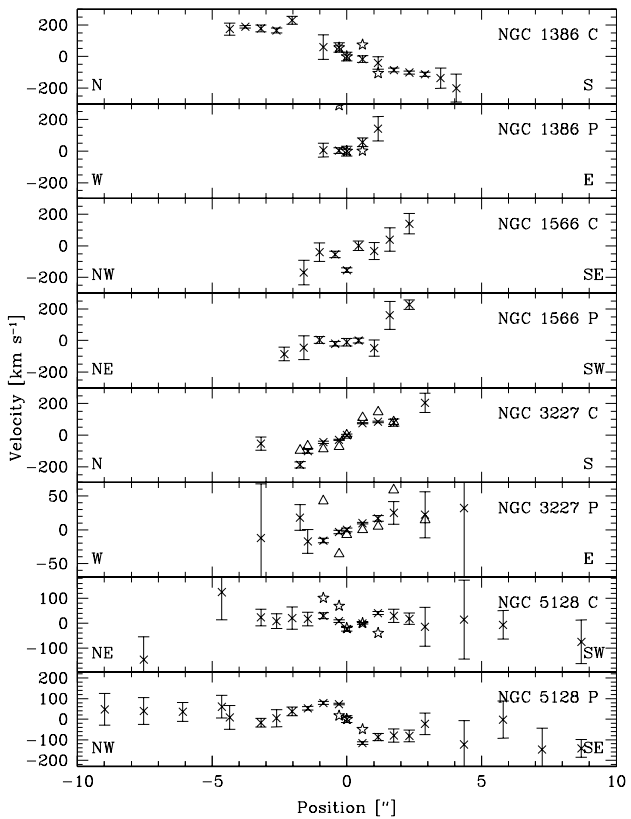,width=8.5cm}
 \caption{The 1--0 S(1) (crosses with errorbars), {\iron} (triangles) and 
{\brg} (stars) velocity curves for NGC 1386, NGC 1566, NGC 3227 and NGC 5128 
parallel to the cone (C) and perpendicular to it (P)\label{velplot}}
\end{figure}

The velocity field of NGC 1386 is displayed in Fig. \ref{velplot}. 
The {\h2} rotation curve is relatively ordered parallel to the cone, 
suggesting that circular motions dominate the kinematics. 

\subsubsection{Two extended regions parallel to the cone}

The most interesting features in 
Fig. \ref{n1386both} are the two extended {\h2} emission regions parallel to 
the cone. Region A covers the area 4\farcs4 -- 2\farcs3 to the north and 
region B 1\farcs1--3\farcs2 to the south of the nucleus. Neither of the 
regions is visible in the broad--band images.
Extinction derived from continuum colours is ${\ak}$ = 0.03 and 
${\ak}$ = 0.09 mag, while that derived from the 1--0 Q(3)/1--0 S(1) ratio is 
${\ak}$ = 0.04 and ${\ak}$ = 0.20. for regions A and B, respectively.
The {\h2} lines in these two regions are narrower than in the nucleus 
(Table \ref{reg1386}), in contrast to the situation in NGC 4945 (Section 
\ref{sect4945}).

\begin{table*}
 \caption{Observed fluxes, reddening-corrected fluxes, gaussian FWHMs and 
equivalent widths for the various {\h2} emission lines in the two extended 
regions in NGC 1386}
 \label{reg1386}
 \begin{tabular}{lllllllllllllllllllllllll}
  \hline
  & \multicolumn{4}{l}{Region A}&\multicolumn{4}{l}{Region B}\\
  Line     & observed &corrected&FWHM & EW  &observed &corrected&FWHM  &EW   \\
           & \multicolumn{2}{l}{10$^{-16}$ ergs s$^{-1}$ cm$^{-2}$} & \AA & \AA &\multicolumn{2}{l}{10$^{-16}$ ergs s$^{-1}$ cm$^{-2}$} & \AA & \AA\\
  \\
  1-0 S(2) &7.0\PM01.0&7.3 &25.6\PM3.2   &3.0\PM0.5 &14.6\PM1.0  &16.1 &23.8\PM1.2    &2.8\PM0.2 \\
  1-0 S(1) &19.5\PM0.8&20.1&24.6\PM1.0   &8.8\PM0.4 &35.7\PM0.8  &39.0 &26.7\PM0.6    &7.3\PM0.2 \\
  1-0 S(0) &4.2\PM0.7 &4.3 &20.4\PM3.9   &2.2\PM0.4 &6.8\PM0.7   &7.4  &21.3\PM2.3    &1.5\PM0.2 \\
  2-1 S(1) & $<$2.3   &$<$2.4&--         &$<$1.2    &$<$2.4      &$<$2.7&--           &$<$0.6\\
  1-0 Q(1) &17\PM2    &18  &22.7\PM1.6   &15\PM2    &29\PM2      &31   &21.9\PM1.4    &10.8\PM0.6 \\
  1-0 Q(3) &134\PM2   &14  &22.8\PM1.5   &12\PM2    &26\PM2      &28   &22.7\PM1.3    &10.1\PM0.5 \\
  \hline 
 \end{tabular}
\end{table*}

Many of the weaker {\h2} lines not detected in the nucleus are visible in the 
two regions due to much fainter continuum in these regions than in the 
nucleus, while the line fluxes are comparable to those in the nucleus. The 
line widths are only slightly larger than the instrumental resolution 
($\sim$ 22 \AA). The 2--1 S(1) line was not detected in 
either region, setting upper limits of $T_{\rmn{vib}}$ as 2600 K and 2100 K 
for regions A and B, respectively. Furthermore, the non-detection of {\brg} 
suggests thermal excitation in shock-heated clouds. The 1--0 S(0)/1--0 S(2) 
ratio is 0.60 and 0.46 for regions A and B, respectively, and is compatible 
with $T_{\rmn{rot}} \simeq $ 1800 K or 3200 K. For region A, this temperature 
is in good agreement with the 2--1 S(1)/1--0 S(1) upper limit 
($T_{\rmn{vib}} < 2100$ K) assuming thermal excitation. 
For region B, the two derived temperatures agree within errors. 

The mass of the excited molecular hydrogen associated with the two regions of 
2\farcs3$\times$0\farcs6 ($\sim$4350 pc$^2$) area is $\sim$10 M$_\odot$ 
($N_{\rmn{H_2}} = 1.9\,10^{17}$ cm$^{-2}$) and $\sim$30 M$_\odot$ 
(3.8 $10^{17}$ cm$^{-2}$), for regions A and B, respectively. These masses are 
comparable to the mass within the 1\farcs1 nuclear aperture, $\sim$20 M$_\odot$.

\subsection{NGC 1566}

\begin{figure*}
 \psfig{figure=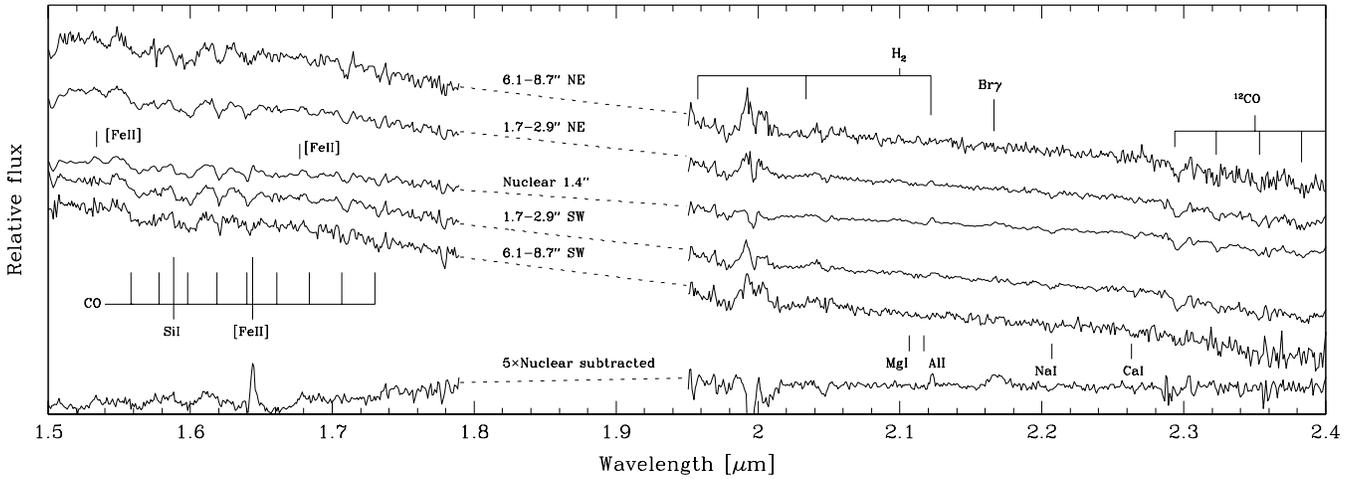}
 \caption{The 1.5--2.5 $\mu$m spectra of NGC 1566 perpendicular to the 
ionization cone.\label{comp1566k}}
\end{figure*}

NGC 1566 is a nearby (distance = 19 Mpc) barred (PA$\sim$ 0$\degr$; Mulchaey, Regan, \& Kundu 1997), 
ringed (diameter 1.7-kpc) SAB(s)bc galaxy in the Dorado group with a Sy1 
nucleus. The nucleus in the high resolution {\oiii} map of 
Schmitt \& Kinney (1996) is almost point-like, with maximum extent of 
0\farcs7 (59 pc). The {\oiii} map also shows conical extended emission in SE. 
To our knowledge, no NIR spectroscopy has previously been 
obtained for NGC 1566. The 1.5--2.5 $\mu$m spectrum is displayed in 
Fig. \ref{comp1566k}.

\begin{figure}
 \psfig{figure=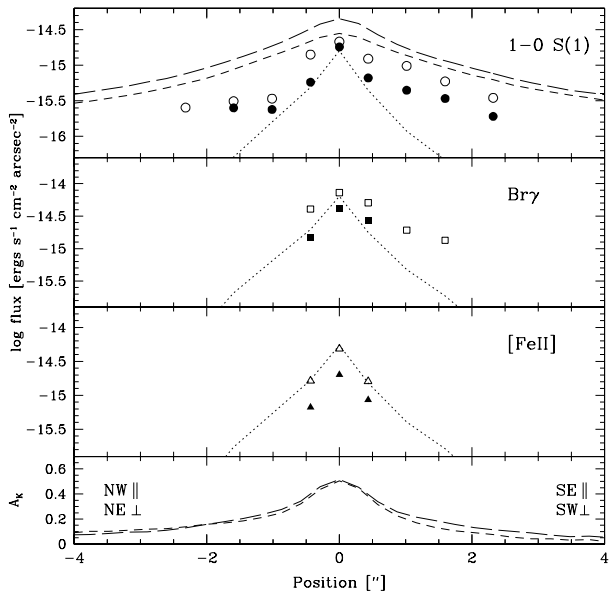,height=8.5cm}
 \caption{The observed line emission in NGC 1566. The panels and symbols are 
as in Fig. \ref{n1386both}, except the continuum emission is multiplied by a 
factor of 5.\label{n1566coneperp}}
\end{figure}

Both the line and the continuum emission appear much weaker parallel to the 
cone than perpendicular to it (Fig. \ref{n1566coneperp}). Of the lines 
detected in NGC 1566 neither {\iron} nor {\brg} are spatially resolved. The 
1--0 S(1) emission is detected up to $\sim$2{\arcsec} ($\sim$200 pc in the plane of the galaxy) from the nucleus 
parallel to the cone and $\sim$ 280 pc perpendicular to it. 
The intrinsic spatial FWHM of the nuclear 1--0 S(1) emission is $\sim$1\farcs3 
perpendicular to the cone and $\sim$1\farcs2 parallel to the cone. The 
1--0 S(1) and {\iron} lines are rather narrow, with intrinsic FWHMs 
430 km s$^{-1}$ and 600 km s$^{-1}$ in the nuclear 1\farcs4 aperture. The 
width of the 1-0 S(1) line is roughly constant along the slit, with no 
detectable difference between the two PAs.

The {\brg} emission line is broad (Fig. \ref{compbrg}), with FWHM = 2100 km s$^{-1}$ 
in the combined nuclear 1\farcs4 aperture after subtracting the neighbouring absorption 
lines. It is detected, however, only within the central 1{\arcsec}, with no 
extended emission. The nuclear {\iron} line is narrow (FWHM = 600 \kms), while 
there appears to be no narrow component to \brg. Similarly, there is no broad 
component to the {\iron} line, which can be explained if extinction 
{$\ak$} $>$ 3 mag. However, the lack of the narrow {\brg} component indicates 
a different excitation mechanism for iron, possibly X-ray excitation 
outside the BLR. The {\iron} line is not spatially resolved,
in agreement with the X-ray heating models by Krolik \& Lepp (1989), which 
predict heating within the torus on scales of a few pc.

The {\h2} 2--1 S(1) line was not detected in the nucleus of NGC 1566, with 
2--1 S(1)/1--0 S(1) $\leq 0.17$ (corresponding to 
$T_{\rmn{vib}} \leq 2700$ K), significantly less than predicted for UV 
fluorescence. The 1-0 S(1)/{\brg} ratio is 0.27, suggesting thermal UV 
heating. The {\iron}/{\brg} ratio is 0.50, but since {\brg} does not display 
a narrow component, a starburst contribution to the UV excitation cannot be 
larger than $\sim$10\%. Therefore, it seems that the dominant mechanism for 
the excitation of the {\h2} gas is collisional via shocks. 
The integrated mass the of excited molecular hydrogen along the direction of the 
ionization cone is $\sim$90 M$_\odot$ and perpendicular to it 
$\sim$110 M$_\odot$. The 1--0 S(1) velocity field (Fig. \ref{velplot}) 
appears relatively ordered, with increasing velocities with radius, thus 
suggesting that circular motions dominate the kinematics.

\subsection{NGC 3227}

\begin{figure*}
 \psfig{figure=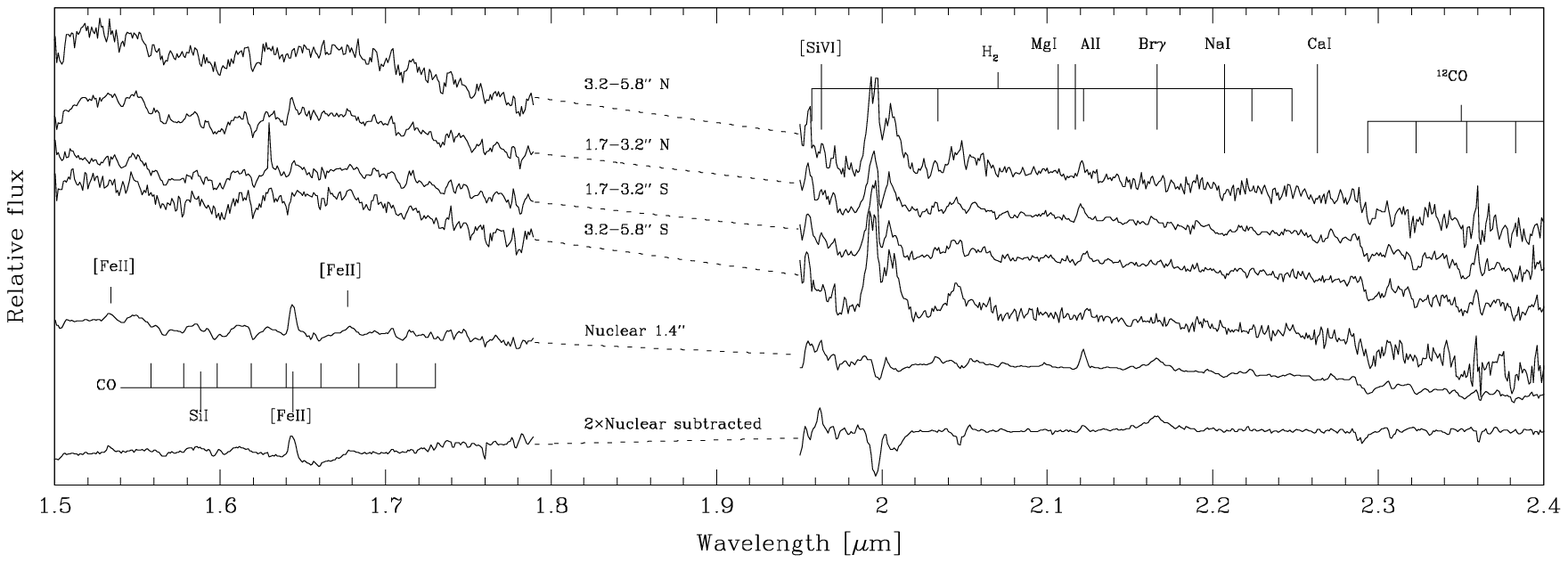}
 \caption{The 1.5--2.5 $\mu$m spectra of NGC 3227\label{comp3227H} parallel to cone}
\end{figure*}

\begin{figure}
 \psfig{figure=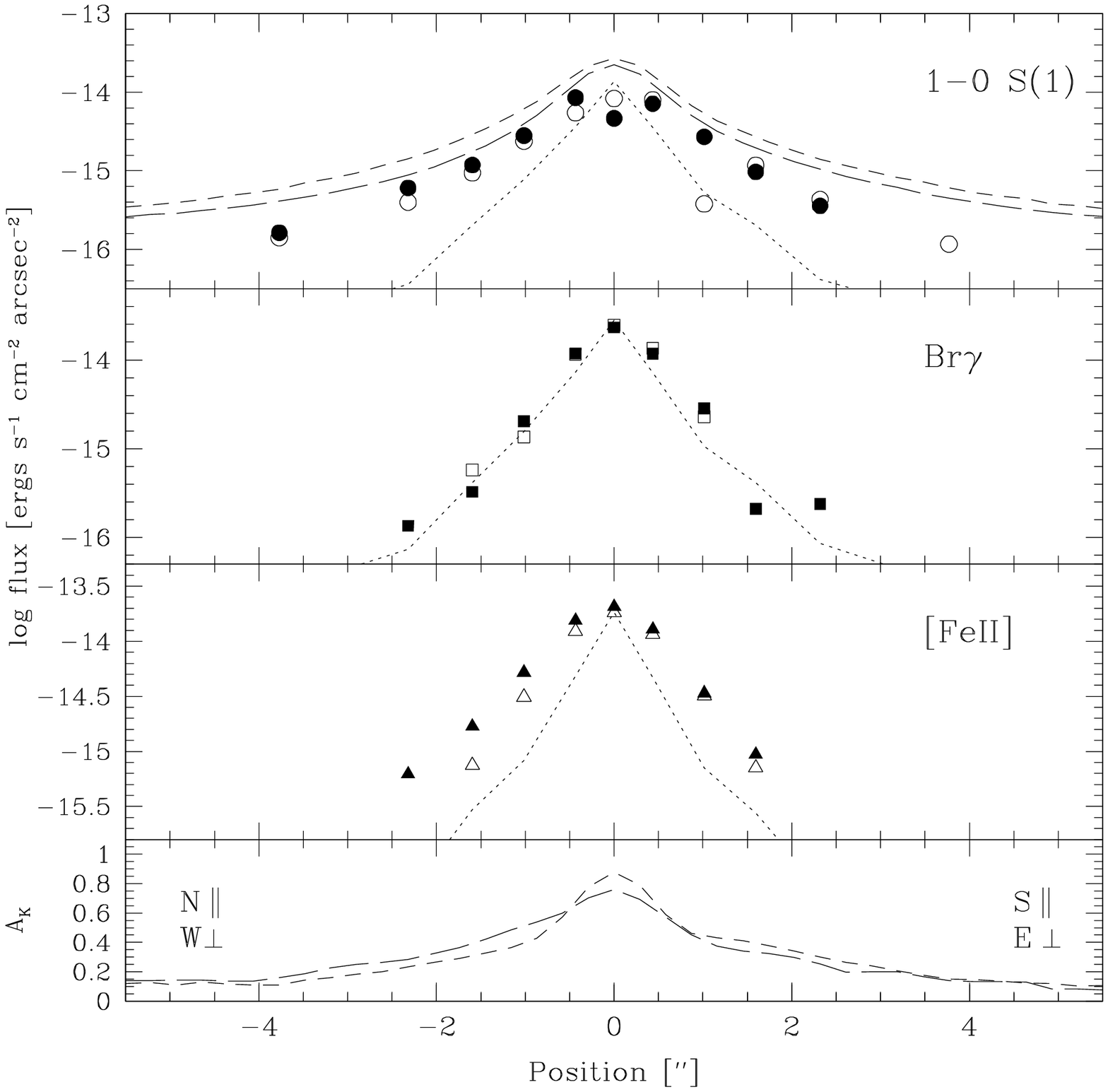,height=8.5cm}
 \caption{The spatial extent of the three main emission lines in NGC 3227. 
For panels and symbols, see Fig. \ref{n1386both}
\label{n3227coneperp}}
\end{figure}

\begin{figure}
 \label{n3227brg}
 \psfig{figure=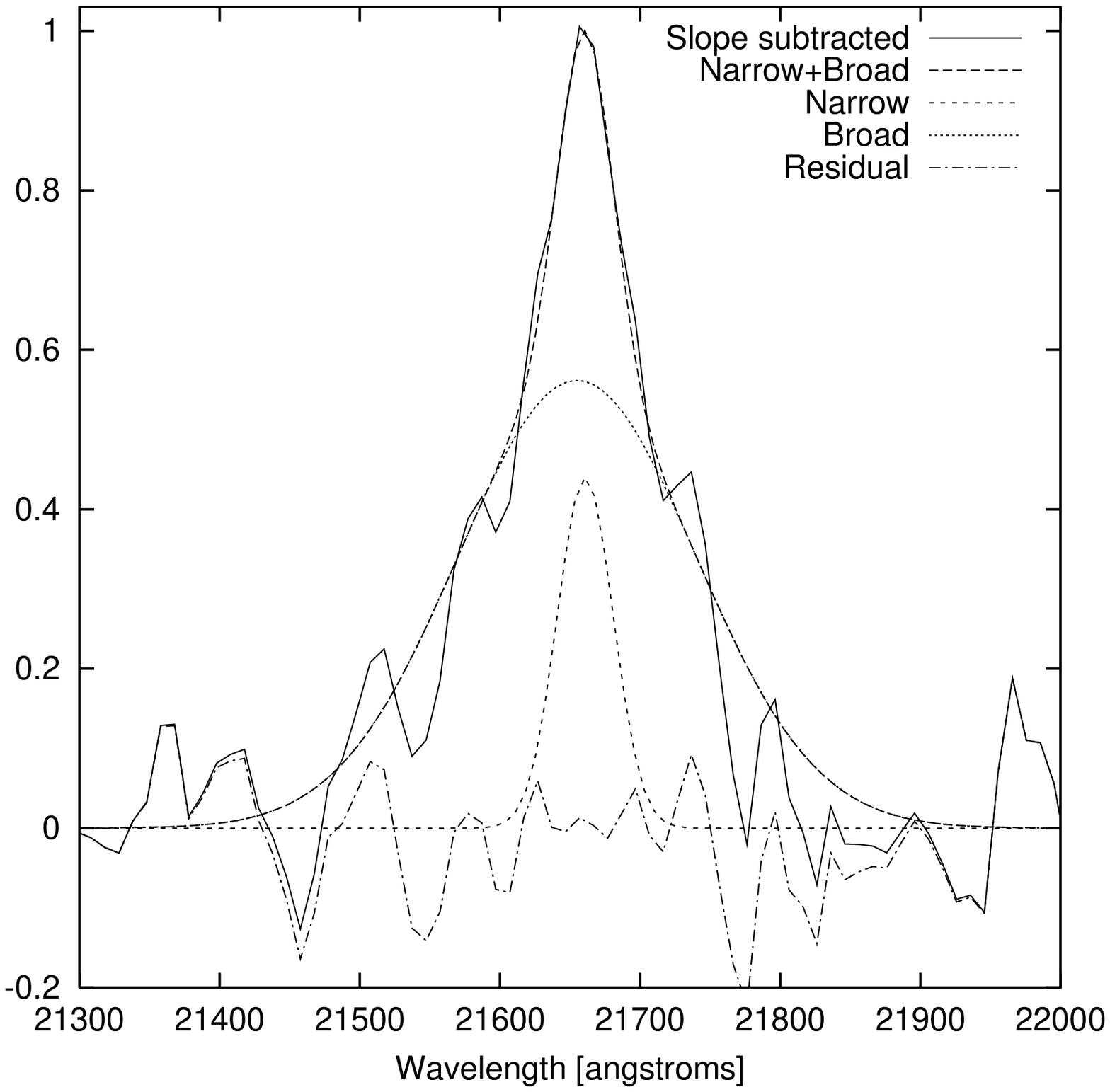,height=8.5cm,width=8.5cm}
 \caption{The {\brg} line profile of NGC 3227. Fit is obtained with a 
narrow component (FWHM = 650 {\kms}, flux 4.2$\times10^{-15}$\ergs), and a 
broad component (FWHM = 2700 \kms, flux 23$\times10^{-15}$\ergs).}
\end{figure}

NGC 3227 is a nearby (distance = 15 Mpc) barred (Mulchaey et al. 1997; PA $\sim 150\degr$) SAB 
galaxy hosting a Sy1.5 nucleus with a one-sided cone to the N (Schmitt \& Kinney 1996).
It has been imaged with HST in the 1--0 S(1) and 1--0 S(3) lines by 
Quillen et al. (1999), although the very strong nuclear emission complicates 
subtracting the continuum in the centre. 
Recently, NIR spectroscopy of NGC 3227 has been obtained by 
Rhee \& Larkin (2000; $R$ = 550) and 
Schinnerer, Eckart \& Tacconi (2001; $R$ = 700-1000). 
In our new spectra the EW of {\iron} and 1--0 S(1), and the integrated 
colours are in good agreement with those derived by Schinnerer et al. 
On the other hand, Br$\gamma$ is much stronger and broader (EW = 8.3 \AA; 
FWHM = 1800 \kms) in our spectra than in those by Schinnerer et al. 
(EW = 3.7 \AA; FWHM = 650 \kms). 

Schinnerer et al. (2001) reported the detection of two coronal lines, 
{\sivi} and {\alix} 2.043 $\mu$m. After subtracting the off-nuclear spectra 
to reduce the residual telluric lines, we found no evidence for the presence 
of {\alix} (Fig. \ref{comp3227H}). The {\sivi} line is present, but at a much 
weaker level than reported by Schinnerer et al. However, this result is 
hardly conclusive because of the atmospheric signatures.

The {\brg} line is broad in NGC 3227 (Fig. \ref{compbrg}), with single 
component FWHM = $\sim$1800 km s$^{-1}$, in agreement with the results by 
Rhee \& Larkin (2000). Unfortunately, outside the nucleus the S/N ratio in 
our spectra is not sufficiently high to accurately subtract the continuum and 
determine the line profile, in the presence of neighbouring absorption lines 
(e.g. 2.1770 $\mu$m and 2.1460 $\mu$m). Fixing the narrow component at FWHM = 
650 {\kms}, as derived by Schinnerer et al. (2001), the FWHM of the broad 
component is 2500 -- 3000 {\kms} (Fig. \ref{n3227brg}), similar to that 
derived by Rhee \& Larkin (2000). 

The nuclear {\brg} emission is spatially unresolved (Fig. \ref{n3227coneperp}), 
with no evidence for extended emission. Therefore, no recent star formation has been 
occurring in the disc of NGC 3227. However, Schinnerer et al. (2001) detected 
a young (25--50 Myr) star cluster close to the nucleus, which may already 
have passed the ionizing stage. Alternatively, it may be the origin of the 
narrow {\brg} component. Neither 1--0 S(1) nor {\iron} have broad 
profiles, indicating that {\brg} originates much closer to the central engine 
than {\iron}. The {\iron} line is broader in the direction of the cone than 
perpendicular to it, and extends predominantly towards N. This effect may be 
due to line splitting, which is not directly detectable with our spectral 
resolution. The narrow {\iron}/narrow {\brg} ratio is $\sim$7.8, which 
suggests that the narrow nebular lines are excited by X-rays, as in the case 
of NGC 1386 or NGC 1566 (see above). Unlike in these galaxies, however, 
{\iron} is spatially resolved in NGC 3227. 

The nuclear 2--1 S(1)/1--0 S(1) ratio is 0.08, clearly ruling out UV 
fluorescence as the dominant excitation mechanism. Outside the nucleus, 
1--0 S(1)/{\brg} ratio $\gg 1$, suggesting shocks to be responsible for the 
excitation of {\h2}. 
In the nuclear 1\farcs4 aperture, the gas density 
$N_{\rmn{H_2}} = 23.4 \, 10^{17}$ cm$^{-2}$, corresponding to gas mass of 
280 M$_\odot$, in agreement with the mass 380 M$_\odot$ derived by 
Fernandez et al. (1999) from the peak {\h2} surface brightness.

Both the 1--0 S(1) emission and, more marginally, the {\iron} emission are 
spatially resolved, with no clear difference in the nuclear extent in the directions 
parallel and perpendicular to the cone. In the 1--0 S(1) image by Quillen 
et al. (1999) the emission is elongated along PA $\sim$ 100\degr. The 1--0 S(1) 
line is broader parallel to the ionization cone than perpendicular to it, 
similarly to {\iron}. This suggests that both the excited {\h2} and {\iron} 
are emitted from the surface of the hollow ionization cone. 
In Schinnerer et al. (2001), the {\iron} line was not spatially resolved, but 
this may be due to their poorer spatial resolution ($\sim$1\farcs6).

The velocity field is well-ordered and resembles solid rotation, 
especially parallel to cone, and {\iron} appears to have the same dynamics 
as that of 1--0 S(1). Perpendicular to the cone the curve is noisier and flat.

\subsection{NGC 4945}
\label{sect4945}

\begin{figure*}
 \psfig{figure=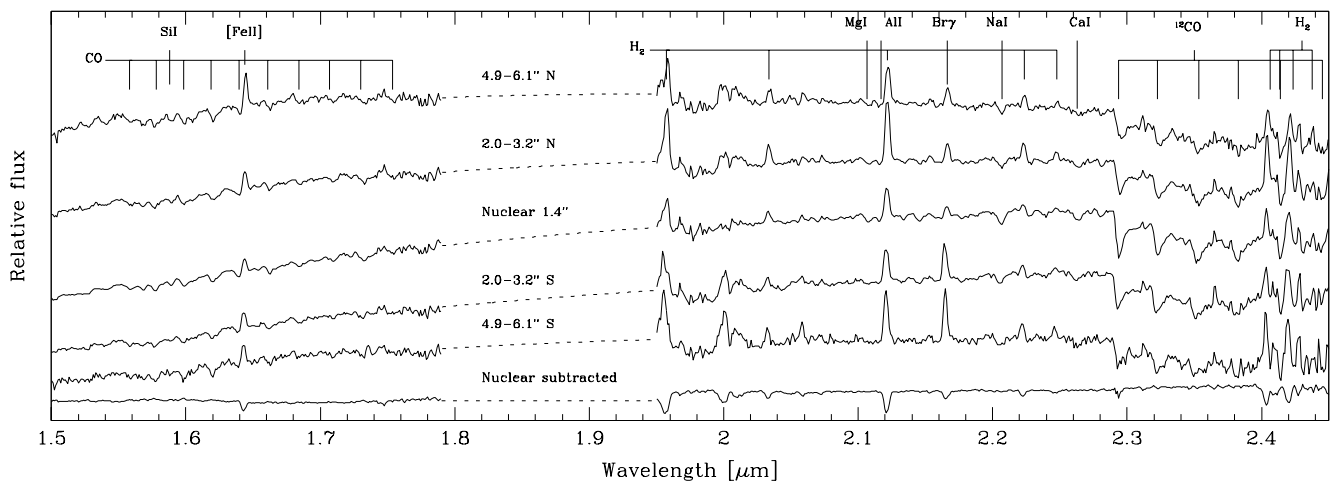}
 \caption{The 1.5 -- 2.5 $\mu$m spectra of NGC 4945 perpendicular to the 
superwind.\label{comp4945K}}
\end{figure*}

NGC 4945 is a nearby SB(s)cd galaxy 
in the Centaurus group of galaxies, with a dust lane SE of the nucleus. 
NGC 4945 has strong starburst activity, concentrated in a circumnuclear ring 
(Marconi et al. 2000) with a projected major axis diameter $\sim$ 11$''$ 
(210 pc). NGC 4945 also harbours a Sy2 nucleus, based on the detection of a 
heavily obscured, rapidly variable, strong hard X-ray source above 10 kev 
(Iwasawa et al. 1993), which is completely obscured below 10 kev. In the 
following we have adopted a distance of 3.9 Mpc for NGC 4945 
(Harris, Poole \& Harris 1998). The hollow cone, which may be 
produced by a superwind (e.g. Heckman, Armus \& Miley 1990) from the powerful 
starburst and not represent an outflow from the nucleus, is well visible in 
shorter wavelengths ($\lambda < 2 \mu$m), especially NW of the nucleus (Moorwood et al. 1996).

\begin{figure}
 \psfig{figure=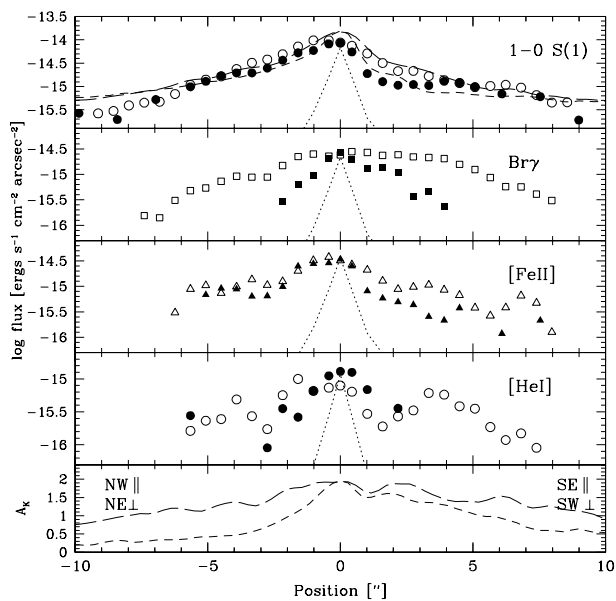,height=8.5cm}
 \caption{The observed line emission in NGC 4945. The panels and symbols are 
as in Fig. \ref{n1386both}.\label{n4945both}}
\end{figure}

Marconi et al. (2000) presented Pa$\alpha$, 1--0 S(1) and {\iron} images of 
NGC 4945. Our new data (Figs. \ref{comp4945K}, \ref{n4945both}) are in good agreement with 
their results. The Pa$\alpha$ image of Marconi et al. shows no central peak 
but is dominated by a star forming ring. 
In the superwind direction, the two {\brg} maxima separated by $\sim$2{\arcsec} 
(180 pc in the plane of the galaxy) correspond to the SE and NW parts of the ring. 
The {\iron} emission in our spectra is also in good agreement with that in 
Marconi et al. The NW peak at $\sim$5{\arcsec} (440 pc) from the nucleus appears as 
an elongation in their image. The {\iron} emission can be traced further out 
toward NW than SE. Another tracer of young stars, {\he} can be traced 
$\sim$6--8{\arcsec} ($\sim$230 pc) from the nucleus perpendicular and 
$\sim$3{\arcsec} (260 pc) parallel to the superwind.
The 1--0 S(1) map in Marconi et al. shows an elongated (PA$\sim$ 33\degr) 
nuclear emission and extended emission to SW and NW of the nucleus, in 
agreement with our spectra. 
Toward SW, the emission rapidly decreases with local peaks at 4{\arcsec} (130 pc) and 
7{\arcsec} (230 pc) distance, while toward NE the decline is slower. 

\begin{figure}
  \psfig{figure=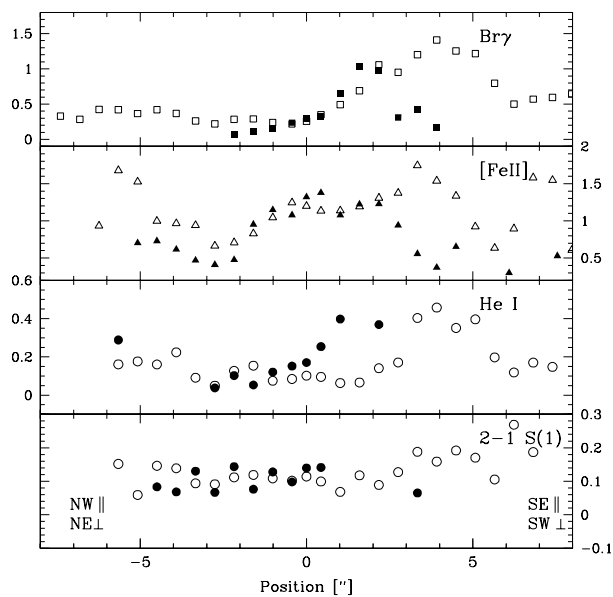,height=8.5cm}
  \caption{The line ratios of the main emission lines with respect to 
1--0 S(1) in NGC 4945.\label{lr4945both}}
\end{figure}

There are quite large radial changes in the line ratios of the main emission 
lines relative to 1--0 S(1) (Fig. \ref{lr4945both}). Notably, the star 
forming region SW of the nucleus perpendicular to the superwind is clearly 
visible. There is a reasonable correlation between {\brg} and {\he}, while 
the behaviour of {\iron} emission is different from them. The 
{\brg}/1--0 S(1) ratio is 0.1--0.5 toward NE and NW of the nucleus, 
suggesting that UV heating is not important. However, toward SE this ratio is 
$\sim$1.0 at 2{\arcsec} (180 pc) distance from the nucleus and toward SW it is 
$\sim$1.4 at 5{\arcsec} (160 pc) distance from the nucleus. These ratios 
indicate, together with the enhanced {\he} and {\iron} emission, that UV 
emission from young, hot stars is important.

The 2--1 S(1)/1--0 S(1) ratio of $\sim$0.1--0.2 (Fig. \ref{lr4945both}) is in 
good agreement with collisional excitation at 
$T_{\rmn{vib}} = 2000 - 2500$ K. The 1--0 Q(1)/1--0 Q(3) ratio in the nucleus 
is $\sim$1, and the associated rotational temperature 
$T_{\rmn{rot}}\simeq 2000$ K. Adopting $T$ = 2000 K, the density of excited 
{\h2} in the nucleus $N_{\rmn{H_2}} \simeq 100\times10^{17}$ cm$^{-2}$ 
(corresponding to $m$ $\simeq$ 70 M$_\odot$), much lower than the available 
mass of warm ($T=160$ K) {\h2} (2.4$\times$10$^7$ M$_\odot$; 
Spoon et al. 2000).

The {\iron}/{\brg} ratio of $\sim4.5$ at the nucleus (Fig. \ref{lr4945both}) 
is lower than expected for X-ray excitation and can better be explained by 
shocks and star formation. Perpendicular to the superwind, the 
\he/{\brg} ratio is roughly constant, with slightly larger ratios ($\sim$0.4) 
toward NE than SW. However, $\sim$1--3{\arcsec} SW there is a region with 
significantly lower ratios ($\sim$0.1--0.2). This may be due to a lower 
number of high mass stars as a result of a decaying starburst or a truncated 
IMF in this region. Parallel to the superwind, this ratio is higher, 
$\sim 0.8$, both 1{\arcsec} NW and 0\farcs5 SE of the nucleus. 

Both the 1--0 S(1) and especially the {\iron}  are significantly broader in the 
direction of the superwind (up to 550 km s$^{-1}$ and 700--800 km s$^{-1}$, respectively) than 
perpendicular to it ($\sim$ 400 km s$^{-1}$) at $\sim$5{\arcsec} (440 pc) distance 
from the nucleus. This is in agreement 
with that observed in {\ha} (Heckman et al. 1990). However, {\iron} 
and 1--0 S(1) are also broader SE of the nucleus unlike {\ha}, probably due 
to lower extinction which in NIR makes it possible to observe the cone 
surface furthest from us. 

\begin{figure}
  \psfig{figure=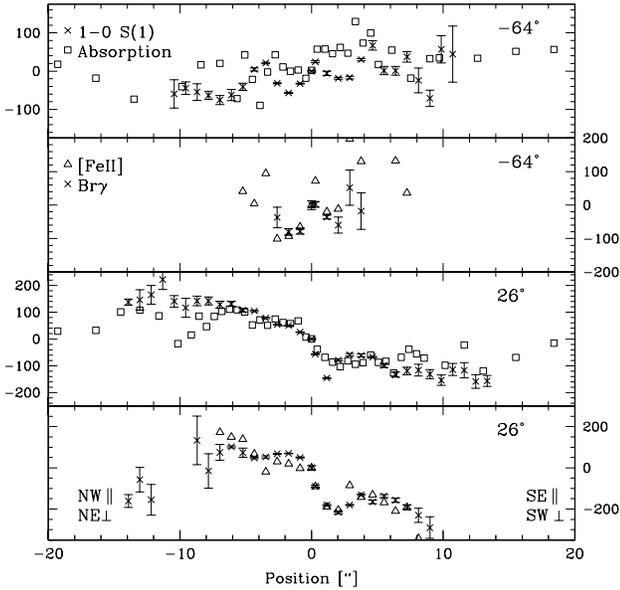,height=8.5cm}
  \caption{The velocity field of the main emission lines and the $K$-band CO-absorption 
lines in NGC 4945 parallel to the superwind (-64\degr) and perpendicular to it (26\degr) 
in the units of km s$^{-1}$. 
\label{n4945vel} }
\end{figure}

The 1--0 S(1) emission has a red wing within a few central arcsec, but it is 
symmetric further out. On the other hand, SW of the nucleus, the red wing in 
both {\brg} and {\iron} is much stronger than the blue wing,
while NE of the nucleus the blue wing is stronger. A likely explanation for 
this behaviour is that the line-of-sight rotational velocity in the nuclear 
ring is higher than in the underlying galaxy, either intrinsically or as a 
result of the ring inclination with respect to the galaxy.

Both perpendicular and parallel to the superwind, the velocity field 
(Fig. \ref{n4945vel}) of 1--0 S(1) is clearly different from that of {\iron} 
and {\brg}, suggesting differences in the dynamics and/or excitation 
mechanism. The kinematics of {\brg} and {\iron} are better correlated, except 
toward NE of the nucleus. 1--0 S(1) follows better the stellar rotation as derived 
from absorption lines than from either {\brg} or {\iron}, but correlation is not 
perfect. The most intriguing feature is the roughly 
sinusoidal form of the rotation curve in the direction of the superwind. It 
is most clearly seen in the various {\h2} lines, but is also present in 
{\brg} and {\he}. Based on the velocity field of {\iron} and assuming 
$i = 78\degr\pm2\degr$ and position angle for the line of nodes $PA = 43\degr\pm1\degr$ (Ables et al. 1987), we 
derive a dynamical mass $M_{\rmn{dyn}} = 3.2 \times 10^{8} {\rmn{M}} _\odot$. 
This value is slightly less than that derived in the inner 
$R \leq 100 {\rmn{pc}}$ from millimeter CO observations 
($M_{\rmn{dyn}}\sim 8 \times 10^8 {\rmn{M}}_\odot$; Mauersberger et al. 1996).

\subsection{NGC 5128}

\begin{figure*}
 \psfig{figure=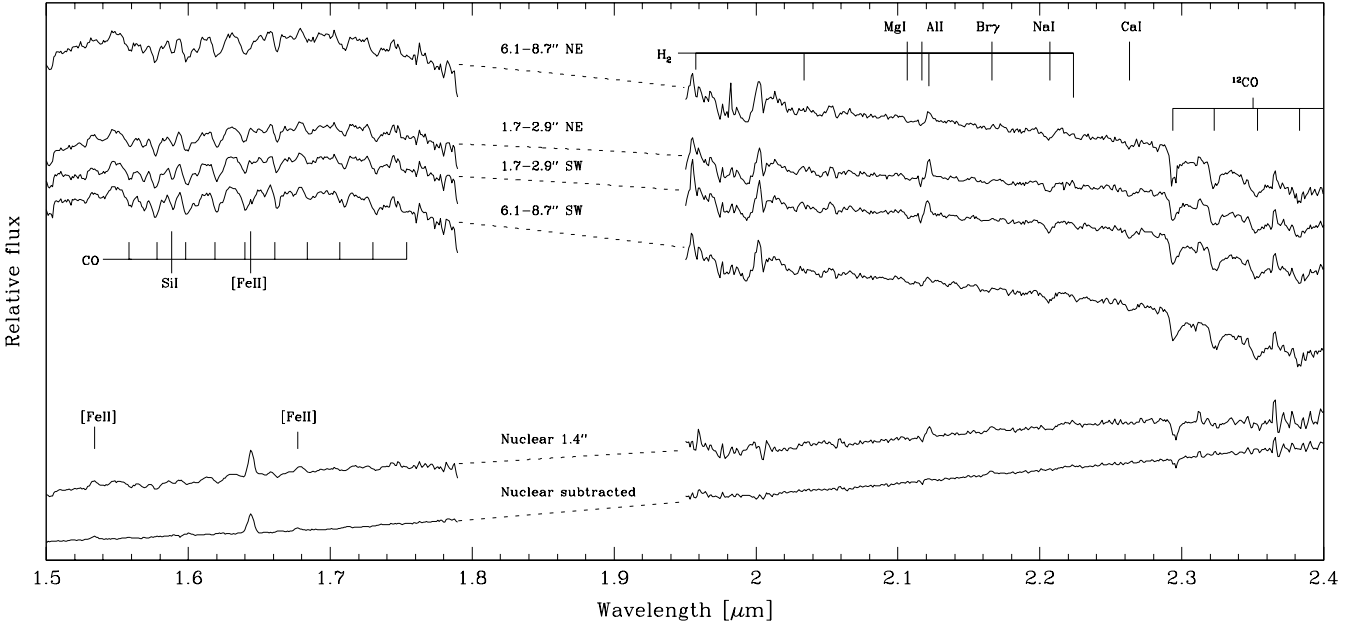}
 \caption{The 1.5--2.5 $\mu$m spectra of NGC 5128 perpendicular to the 
jet.\label{comp5128K}}
\end{figure*}

NGC 5128 (Centaurus A) is a well-studied galaxy with prominent dust 
lanes (e.g. Schreier et al. 1996) and Sy2 nucleus. It has very prominent 
radio jets which after 5-kpc expand into plumes. 
The distance of NGC 5128 has been controversial, with estimates 
ranging from 2.1 Mpc to 8.5 Mpc; in the following, we have adopted the 
distance of 3.9 Mpc (Harris et al. 1998). Previous NIR spectroscopic studies of NGC 5128 include e.g. Marconi et al. (2001) and Bryant \& Hunstead (1999). In addition, it has 
been the target of NIR Pa$\alpha$ (Schreier et al. 1998) and {\iron} 
(Marconi et al. 2001) imaging. 
For a comprehensive review of NGC 5128, see Israel (1998).

The spectra of NGC 5128 at different distances from the nucleus perpendicular 
to the jet are displayed in Fig. \ref{comp5128K}. The subtracted nuclear 
spectrum is almost featureless. There are some notable differences between 
our spectra and those from the literature. Especially, while the 
nuclear 1\farcs4$\times$1\farcs6 {\iron}/{\brg} ratio observed by 
Simpson \& Meadows (1998) = 1.47,
we find this ratio to be 5.1, and after correction for extinction even 
$\sim$15, one of the largest ratios ever found. 

\begin{figure}
 \psfig{figure=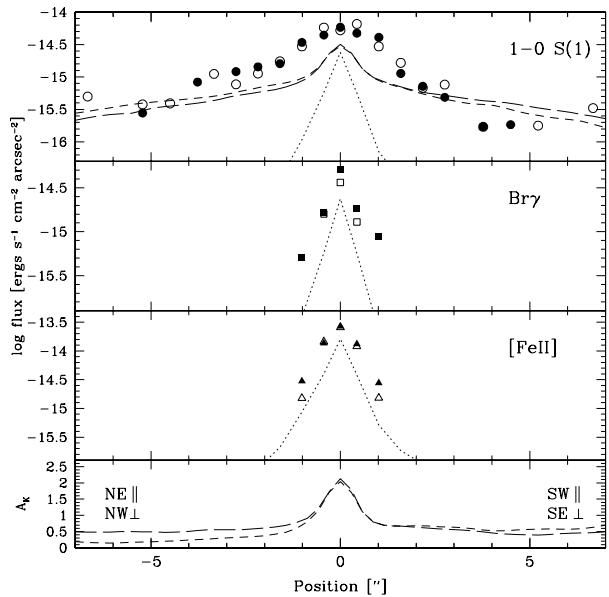,height=8.5cm}
 \caption{The observed line emission in NGC 5128. Symbols are as in 
Fig. \ref{n1386both} \label{n5128coneperps1} }
\end{figure}

The most striking feature in the spectra of NGC 5128 is the weakness of the 
line emission compared to that in NGC 4945 at similar distance. This is 
partly due to dilution by the bright, red, featureless nuclear continuum.
{\iron} is the strongest line in the nuclear region. However, the only 
emission line that can be traced out into the galaxy is 1--0 S(1) 
(Fig. \ref{n5128coneperps1}), which extends parallel to the jet up to 
13\farcs4 (320 pc in the plane of the galaxy) in NE and 7\farcs4 (180 pc) in 
SW, and perpendicular to the jet up to 16{\arcsec} (690 pc) in NW and 
22{\arcsec} (950 pc) in SE. The nuclear {\brg} emission is not spatially 
resolved. In the deeper $K$-band spectra of Marconi et al. (2001) the nucleus 
is resolved and detected up to 1{\arcsec} from the nucleus. Marconi et al. 
interpreted this as an inclined, $\sim$40 pc diameter, thin disc of ionized 
gas. In addition to central component, there are a few star forming regions 
parallel to the jet visible in our spectra, e.g. at $\sim$17{\arcsec} (400 pc) 
with detected {\brg} and {\iron} emission. This is in agreement with the 
Pa$\alpha$ and {\iron} images of Schreier et al (1998) and Marconi et al. 
(2001), respectively, which display two extended regions straddling the 
nucleus at PA $\sim$33\degr. The {\iron}-emission of Marconi et al. is mostly 
unresolved with faint extended emission NW of the nucleus.

No 2--1 S(1) emission was detected in the nucleus, with an upper limit of 
2--1 S(1)/1--0 S(1) $\leq 0.08$ ($T_{\rmn{vib}}<$ 1900 K). Since the 
1--0 S(1)/{\brg} ratio is 2.04, the {\h2} gas is probably excited by shocks. 
Assuming $T_{\rmn{vib}}=2000 K$, the density of {\h2} is 
45.6$\times10^{17}$ cm$^{-2}$, corresponding to a gas mass of 33.8 M$_\odot$.

Our velocity field for NGC 5128 is displayed in Fig. \ref{velplot}. It is in 
agreement with Marconi et al. (2001), but unfortunately we lack the necessary 
resolution or depth to supplement their results. 

\section{Conclusions}

We have presented NIR 1.5--2.5 $\mu$m long-slit spectra of six Seyfert 
galaxies with an ionization cone or jets. While statistical analysis will be 
presented in forthcoming papers using the full sample of 14 Seyferts, some 
general trends can already be noted. {\iron} is the strongest and {\brg} the 
weakest main emission line in the Sy2s NGC 1386, NGC 4945 and NGC 5128.
On the other hand, in the Sy1 NGC 1097 1--0 S(1), and in the Sy1 NGC 1566 and 
the Sy1.5 NGC 3227 {\brg} is the strongest emission line. {\brg} was not 
detected in the nucleus of NGC 1097. Notably, in NGC 1386 the coronal {\sivii} 
line is even stronger than {\iron}, {\sivi} is comparable to {\iron} and 
{\caviii} is stronger than {\brg}. In addition, {\sivi} and {\sivii} were 
also detected in NGC 3227.

Broad nuclear {\brg} line emission was detected in NGC 1386, NGC 1566 and 
NGC 3227, in which narrow {\brg} component is only visible in NGC 3227. 
Based on the high {\iron}/{\brg}-ratios it seems probable that {\iron} is 
X-ray excited. The {\iron} emission is spatially resolved only in 
NGC 3227 and NGC 4945, which also agrees with the theory of X-ray excitation 
in NLR as the origin for the {\iron} emission (e.g. Mouri, Kawara \& Taniguchi 2000).

The 1--0 S(1) emission is extended in all galaxies, except parallel to the 
cone in NGC 1386. The extended 1--0 S(1) emission can be detected up to a 
range of physical scales, from $\sim200$ pc in the plane of the galaxy in NGC 
1097 to $\sim$1.4-kpc in NGC 4945. In NGC 1097, NGC 1386 and NGC 4945 separate 
off-nuclear 1--0 S(1) emission regions are detected. In NGC 1097 and NGC 4945 
these regions are associated with the starburst rings, while in NGC 1386 the 
origin of these regions remains unknown, but are likely to be connected with 
a nuclear outflow. Since the extended emission declines smoothly with 
increasing radius, it seems probable that the molecular gas forms a disc 
surrounding the nucleus. This is further supported by 1--0 S(1) velocity 
curves, which are generally well-ordered except parallel to the superwind in 
NGC 4945. A starburst origin seems excluded as extended {\brg} or {\iron} is 
only detected in NGC 1097 and NGC 4945. In these galaxies, the extended 
emission is associated with off-nuclear star forming regions

The overall 1--0 S(1) emission is more extended parallel to the cone/jet than 
perpendicular to it in NGC 1097, NGC 1386  and NGC 4945, while the opposite 
is true in NGC 3227, NGC 1566 and NGC 5128, so the total extent of the 
emission does not depend on the type of the nucleus. The extent of the 
nuclear core {\h2} emission is larger perpendicular to the ionization 
cone/jet than parallel to it in NGC 1097, NGC 1386, NGC 1566, NGC 3227 
and NGC 4945, in good agreement with the current unified models and the 
existence of a molecular torus. In NGC 5128, on the other hand, the extent is 
larger parallel to the jet than perpendicular to it. 

There does not appear to be a clear difference between the gas mass 
M$_{{H}_2}$ parallel and perpendicular to the ionization cone. In NGC 1566, 
NGC 3227, and especially in NGC 4945, M$_{{H}_2}$ parallel to the cone is 
smaller than perpendicular to it. On the other hand, in NGC 5128 the two 
masses are similar, while in NGC 1097 and NGC 1386 M$_{{H}_2}$ parallel to 
the cone/jet is larger than perpendicular to it. The highest nuclear gas 
densities $N_{\rmn{H_2}}$ are found in the Sy2s NGC 4945 and NGC 5128, and 
the lowest in the Sy1s NGC 1097 and NGC 1566. In the Sy1.5 NGC 3227 the 
density is intermediate, while in the Sy2 NGC 1386 the density is comparable 
to that in the Sy1s. 

In all galaxies, the dominant excitation mechanism of the nuclear {\h2} 
emission appears to be due to shocks. No evidence for significant UV 
fluorescence was found. The extended {\h2} emission in NGC 1097, NGC 1386 and 
NGC 4945 is also best explained by thermal excitation. 

\section{Acknowledgements}
This research has made use of the NASA/IPAC Extragalactic Database (NED) 
which is operated by the Jet Propulsion Laboratory, California Institute of 
Technology, under contract with the National Aeronautics and Space 
Administration. 
\\

\noindent{\bf References:}\\

\noindent
Ables J.G., Forster J.R., Manchester R.N. et al., 1987, MNRAS, 226, 157\\
Alonso-Herrero A., Rieke M.J., Rieke G.H., Ruiz M., 1997, ApJ, 482, 747\\
Antonucci R., 1993, ARA\&A, 31, 473\\
Bryant J.J., Hunstead R.W., 1999, MNRAS, 308, 431\\
Contini M., Viegas S.M., 2001, ApJS, 132, 211\\
Fernandez B.R., Holloway A.J., Meaburn J., Pedlar A., Mundell C.G., 1999, MNRAS, 305, 319\\
Fioc M., Rocca-Volmerange B., 1999, A\&A, 351, 869\\
Gredel R., Dalgarno A., 1995, ApJ, 446, 852\\
Harris G.L.H., Poole G.R., Harris W.E., 1998, AJ, 116, 2866\\
Heckman T.M., Armus L., Miley G.K., 1990, ApJS, 74, 833\\
Hollenbach D., McKee C.F., 1989, ApJ, 342, 306\\
Hunt L., Malkan M.A., Salvati M., et al., 1997, ApJS, 108, 229\\ 
Israel F.P., 1998, A\&AR, 8, 237\\
Iwasawa K., Koyama K., Awaki H., et al., 1993, ApJ, 409, 155\\
Kotilainen J.K., Reunanen J., Laine S., Ryder S., 2000, A\&A, 353, 834\\
Krolik J.H., Lepp S., 1989, ApJ, 347, 179\\
Landini M., Natta A., Salinari P., Oliva E., Moorwood A.F.M., 1984, A\&A, 134, 284\\
Lidman C., Cuby J.-G., Vanzi L., 2000 SOFI User's manual, LSO-MAN-ESO-40100-0003\\
Maiolino R., Ruiz M., Rieke G.H., Papadopoulos P., 1997, ApJ 485, 552\\
Malkan M.A., Gorjian V., Tam R., 1998, ApJS, 117, 25\\
Maloney P.R., Hollenbach D.J., Tielens G.G.M., 1996, ApJ, 466, 561\\
Marconi A., van der Werf P.P., Moorwood A.F.M., Oliva E., 1996, A\&A, 315, 335\\
Marconi A., Schreier E.J., Koekemoer A., et al., 2000, ApJ, 528, 276\\
Marconi A., Capetti A., Axon D., et al., 2001, ApJ, 549, 915\\
Mauersberger R., Henkel C., Whiteoak J.B., Chin Y.-N., Tieftrunk A.R., 1996, A\&A, 309, 705\\
Moorwood A.F.M., Oliva E. 1988, A\&A, 103, 278\\
Moorwood A.F.M., van der Werf P.P, Kotilainen J.K., Marconi A., Oliva E. 1996, A\&A, 308, 1L\\
Moran E.C., Barth A.J., Kay L.E., Filippenko A.V., 2000, ApJ, 540, L73\\
Mouri H., Kawara K., Taniguchi Y., 2000, ApJ, 528, 186\\
Mulchaey J.S., Wilson A.S., Tsvetanov Z., 1996, ApJ, 467, 197\\
Mulchaey J.S., Regan M.W., Kundu A., 1997, ApJS, 110, 299\\
Nagar N.M., Wilson A.S., Mulchaey J.S., Gallimore J.F., 1999, ApJS, 120, 209 \\
Oliva E., Salvati M., Moorwood A.F.M., Marconi A., 1994, A\&A, 288, 457\\
Phillips M.M. Pagel B.E., Edmunds M.G., Diaz A., 1984, MNRAS, 210, 701\\
Quillen A.C., Alonso-Herrero A., Rieke M.J., et al., 1999, ApJ, 527, 696\\
Rhee J.H., Larkin J.E., 2000, ApJ, 538, 98\\
Rossa J., Dietrich M., Wagner S.J., 2000, A\&A, 362, 501\\
Schinnerer E., Eckart A. Tacconi L.J., 2001, ApJ, 549, 254\\
Schmitt H.R., Kinney A.I., 1996, ApJ, 463, 498\\
Schreier E.J., Capetti A., Macchetto F., Sparks W.B., Ford H.J., 1996, ApJ, 459, 535\\ 
Schreier E.J., Marconi A., Axon D., et al. 1998, ApJ, 499, L143\\
Scoville N.Z., Hall D.N.B., Kleinmann S.G., Ridgway S.T., 1982, ApJ, 253, 136\\
Simpson C., Meadows V., 1998, ApJ, 505, L99\\
Spoon H.W.W., Koornneef J., Moorwood A.F.M., Lutz D., Tielens A.G.G.M., 2000, A\&A, 357, 898\\
Sternberg A., Dalgarno A., 1989, ApJ, 338, 197\\
Storchi-Bergmann T., Baldwin J.A., Wilson A.S., 1993, ApJ, 410, 11\\
Storchi-Bergmann T., Wilson A.S., Baldwin J.A., 1996a, ApJ, 460, 252\\
Storchi-Bergmann T., Rodr\'\i guez-Ardila A, Schmitt H.R., Wilson A.S., Baldwin J.A., 1996b, ApJ, 472, 83\\
Storchi-Bergmann T., Winge C., Ward M.J., Wilson A.S., 1999, MNRAS, 304, 35\\
Turner J., Kirby-Docken K., Dalgarno A., 1977, ApJS, 35, 281\\
Veilleux S., Goodrich R.W., Hill G.J., 1997, ApJ, 477, 631\\
Weaver K.A., Wilson A.S., Baldwin J.A., 1991, ApJ, 366, 50\\
Wehrle A.E., Keel W.C., Jones D.L., 1997, AJ, 114, 115\\
Winge C., Storchi-Bergmann T., Ward M.J., Wilson A.S., 2000, MNRAS, 316, 1\\
Wolstencroft R.D., Tully R.B., Perley R.A., 1984, MNRAS, 207, 889\\

\label{lastpage}
\end{document}